\documentclass[journal]{IEEEtran}

\usepackage{amsmath,amsfonts,mathrsfs}
\usepackage{array}
\usepackage{textcomp}

\usepackage{url}
\usepackage{verbatim}
\usepackage{graphicx}
\usepackage{subfigure}
\usepackage{stfloats}
\usepackage{cite}
\usepackage{bm}

\usepackage{booktabs}

\usepackage{hyperref}

\begin{document}

\title{Unlocking Cross-Scenario Physical Layer Security: A Mixture-of-Experts Framework with\\Generative Diffusion Models}

\author{Xiao Tang, Tong Hui, Chao Shen, Yichen Wang, Qinghe Du, Li Sun, and Zhu Han%
\thanks{X. Tang is with the School of Information and Communication Engineering, Xi'an Jiaotong University, Xi'an 710049, China, and also with Shenzhen Research Institute of Northwestern Polytechnical University, Shenzhen 518057, China. (e-mail: tangxiao@xjtu.edu.cn)}
\thanks{T. Hui is with the School of Electronics and Information, Northwestern Polytechinical University, Xi'an 710072, China.}
\thanks{C. Shen is with the Faculty of Electronics and Information Engineering, Xi’an Jiaotong University, Xi’an 710049, China.}
\thanks{Y. Wang and Q. Du are with the School of Information and Communication Engineering, Xi'an Jiaotong University, Xi'an 710049, China.}
\thanks{L. Sun is with Department of Network Intelligence, Pengcheng Laboratory, Shenzhen 518055, China.}
\thanks{Z. Han is with the Department of Electrical and Computer Engineering, University of Houston, Houston 77004, USA.}
}



\maketitle

\begin{abstract}
The future 6G networks are expected to incorporate a proliferation of wireless services in diverse environments, which presents a significant challenge for information security. Conventionally optimization always requires recalculation and learning strategy often suffers poor generalization, which are thus incapable for the security provisioning with wide scenario coverage. In this paper, we propose an adaptive and robust learning framework that leverages a mixture-of-experts (MoE) architecture to achieve cross-scenario physical layer security guarantee. Specifically, we first select a few representative scenarios and establish the scenario-specific generative diffusion model (GDM)-based experts for secure transmission beamforming with artificial noise. The diffusion nature of experts learns the overall probability distribution of security strategy solution landscape and the Transformer-based denoising process enhances the ability to generalize across varying network configurations. Then, a lightweight gating network is constructed to identify the scenarios by engineering the channel features and select the most relevant experts. Finally, an attention-based combiner is introduced to synthesize the security proposals from the top-rated experts to produce a high-fidelity security strategy to cover the unseen scenarios. Simulation results demonstrate that the proposed GDM-based MoE framework can accurately recognize the scenarios and properly select the experts, maintaining near-optimal secrecy rates across a continuum of wireless scenarios and outperforming traditional single-model paradigms.
\end{abstract}


\section{Introduction}

Information security is the fundamental and pervasive requirement for wireless communications, which is yet challenged by various malicious adversaries. The rapid development of 6G networks has brought a proliferation of wireless devices and services, which demand reliable and diverse security provisioning~\cite{6g}. Despite the indispensability of upper-layer cryptography-based approaches, the overhead induced by the key management and distribution can be prohibitive for the vast amount of wireless devices which are usually of limited resources and capabilities~\cite{6g-key}. Consequently, the physical layer security with keyless operations facilitates a powerful complement, enabling the lower-layer defense by exploiting the inherent randomness of the wireless medium~\cite{6g-pls}. With secrecy enhancement implemented in the air, the physical layer security is capable for ubiquitous security guarantee across different devices and applications. This advantage aligns with vision of future 6G to cover diverse services and scenarios, making physical layer security a fascinating solution towards a safeguarded 6G era~\cite{6g-pls2}.

For the extensive and continued research efforts on physical layer security, the majority are based on optimizations which are usually burdened with complex iterations and recalculations in changing circumstances~\cite{dl-opt}. Recently, the deep learning-based security designs, including convolutional neural networks, graph neural networks, etc., offer an alternative and promising path forward and thus are emerging rapidly~\cite{dl-l}. Although these approaches feature prompt inferences with trained neural networks, they are fundamentally discriminative in nature that approximate a single-point estimation, making the model potentially brittle for unseen input~\cite{dl-g}. Given the rapidly changing wireless environment, even slightly channel variations not encountered during training may induce significant security performance degradation~\cite{dl-d}. Moreover, the complex relationship between the security strategy and performance leads to potentially vast and complicated space of the (near-)optimal solutions, which is rather difficult to approximate under the discriminative models~\cite{dl-sec}.

Consequently, the recent work resorts to the generative models to overcome such deficiencies, where the generative diffusion model (GDM) is particularly prominent for its high-quality and controllable generation capability~\cite{gdm1,gdm2}. Unlike discriminative models that establish a direct and deterministic mapping, the GDMs learn the entire conditional probability distribution between the wireless environment and the optimal security strategy. In this regard, GDMs can capture the complex structure of the overall solution landscape, producing diverse and high-fidelity candidate solutions to tackle the random fading environments~\cite{gdm-sec}. This capability is critical for physical layer security, where the system performance is challenged by the constant channel fluctuations and unpredictable adversary behavior. In this context, GDMs are promising to produce highly robust and flexible security countermeasures to achieve resilient and adaptive security protection.

When considered further, even a powerful generative model is significantly challenged when confronted with large dynamic of wireless networks, particularly for 6G communications with a variety of emerging scenarios demanding reliable security provisioning~\cite{moe-sce}. In this regard, despite the generative capability, a single GDM is still fundamentally limited by the training data statistics, as a model explicitly trained over a specified data set associated with one scenario can hardly maintain the performance when the scenario characteristics become different~\cite{moe-lim}. For example, the security loss is almost inevitable when a GDM-based security solution initially trained for urban settings is reused in suburban environments. To fill the gap between the inherent limitation of static-model issue and the cross-scenario applications in 6G, a higher-level architecture is imperative to enable coverage to diverse scenarios with adaptive security guarantee~\cite{moe1}.

Towards the orchestrated expertise of specialized models to achieve cross-scenario secure transmissions, we investigate the multi-user multiple-input single-output (MISO) communications in the presence of eavesdroppers under various scenarios, and propose a mixture-of-experts (MoE) framework with GDM-based security solutions. Specifically, the main contributions are summarized as follows:
\begin{itemize}
	\item We consider the multi-user MISO physical layer secure communication system across a continuum of wireless scenarios, where the propagation environments are quantitatively characterized by the path loss exponent and Rician factor. Accordingly, we formulate the problem to develop a unified security solution that is effective across different scenarios, rather than conventional environment-specific approaches.

 	\item To enable scalable cross-scenario security provisioning, we propose a MoE-based architecture that orchestrates a pool of scenario-specialized security experts with scenario-aware adaptation. In this regard, the expert modeling adopts a conditional generative formulation to capture the solution landscape of secure beamforming, and a two-stage training paradigm is developed to combine stable supervised diffusion learning with secrecy-rate-oriented fine-tuning.

	\item We introduce a lightweight router as the higher-level scenario recognizer, where a fixed-dimensional scenario feature is engineered from high-dimensional channel observations. The extracted features keep sensitive to propagation characteristics yet applicable under varying network configurations, based on which the top-rated experts are activated to establish the practical MoE workflow for cross-scenario operation.

	\item To effectively cover the unseen cross-scenarios, we further design an attention-based combiner to process the expert output. The attention mechanism naturally handles a variable number of activated experts, and a residual correction is learned to capture the non-linear mapping from the weighted expert proposals to a high-fidelity security strategy.

	\item Through extensive simulations, we validate that the proposed architecture maintains near-optimal secrecy rates across varying network settings and diverse scenarios. We also demonstrate reliable scenario recognition by the router, and show that the attention-based residual synthesis consistently improves cross-scenario security performance compared with straightforward weighted expert averaging.
\end{itemize}

The remainder of this paper is organized as follows. Sec.~\ref{sec:rw} reviews the related work. Sec.~\ref{sec:sys} details the system model and problem formulation. Sec.~\ref{sec:gdm} presents the individual GDM experts for specified scenarios. Sec.~\ref{sec:moe} describes the design of the MoE framework, including the router and the combiner, on the basis of GDM experts. Sec.~\ref{sec:sim} provides comprehensive simulation results, and Sec.~\ref{sec:con} concludes this paper.

\section{Related Work} \label{sec:rw}

Attracted by the advantages of physical layer security, the research into this issue has been extensive, where primary methods involve the design of security-aware signals and transmission strategies for secrecy enhancement~\cite{dl-opt}. In this regard, the dominant techniques are the secure beamforming and artificial noise (AN) injection, where the former mainly aims to align with the legitimate direction and letter downgrades the unintended reception~\cite{opt2}. Recently, these classical approaches are jointly investigated with the emerging 6G techniques for further security enhancement, such as cooperative relaying~\cite{opt3}, full-duplex transmissions~\cite{opt4}, non-orthogonal multiple access~\cite{opt5}, reconfigurable intelligent surfaces~\cite{opt6}, programmable antenna systems~\cite{opt7}, and aerial space diversity~\cite{opt8}. These security problems and their variants are typically formulated as optimizations, which are commonly solved by using different iterative techniques. Despite the capability to approach the optima, they are usually burdened with computation complexity that becomes significantly more severe with larger network size. Moreover, the same complexity of re-execution is required whenever the network condition becomes different, making them hardly suited for the escalated network dynamics along with 6G communications.

Recently, there have been increasing research efforts in learning-based physical layer security solutions, where a direct mapping between the channel condition and security strategy can be established in a data-driven manner~\cite{dl-l}. In this respect, various models such as multi-layer perceptrons (MLPs)~\cite{dis1}, convolutional neural networks~\cite{dis2}, and graph neural networks~\cite{dis3} have been applied for secure transmission beamforming~\cite{dis4}, physical layer authentication~\cite{dis5}, secrecy key generation~\cite{dis6}, etc. Although the advantage of these approaches with prompt inference for implementation can be rather attractive, their discriminative nature makes them potentially brittle to environmental variations, resulting in possible out-of-distribution non-robustness in practical applications.

A powerful alternative category of learning approach is based on the generative models, which approximate the overall distribution of data~\cite{gen0}. Early attempts have resorted to the generative adversarial networks~\cite{gen1} and variational autoencoders~\cite{gen2} for generative security designs. Recently, the GDMs have emerged as the frontier methods enabling high-quality strategy output, such as the generative secure Internet of Things communications~\cite{gen3}, secure aerial collaborations~\cite{gen4}, and secure wireless sensing~~\cite{gen5}. In this regard, the application of GDMs for wireless secrecy is still unexplored and thus there remains great potential to further reveal. Moreover, one of the central challenge for any specific learning model is the domain generalization issue, as the essential to cover the various scenarios in the 6G networks~\cite{gen6}. Towards this issue, the MoE framework offers a scalable solution, and a few recent studies employ the MoE for semantic communications~\cite{gen7} and wireless resource allocation~\cite{gen8}. Despite the potential, the MoE architecture has not, to the best of our knowledge, been leveraged for physical layer security. Therefore, we in this work establish a MoE-based approach to achieve instantaneous environmental adaptation while taking the advantage of the GDM-powered security expertise, achieving resilient security provisioning across diverse wireless scenarios.

\section{System Model and Problem Formulation} \label{sec:sys}

\subsection{System Model}

We consider a downlink multi-user MISO system where a base station (BS), equipped with $M$ antennas, serves $K$ single-antenna legitimate users, denoted by the set $\mathcal{K} = \{1, \dots, K\}$. Meanwhile, each legitimate receiver is associated with one single-antenna eavesdropper that intends to wiretap the transmitted message. Practically, the associated eavesdropper can be interpreted as an effective worst-case wiretap receiver for the user (e.g., the most threatening one). The BS employs linear precoding to transmit the confidential data symbol $s_k$ to user $k$, with $\mathbb{E}[|s_k|^2] = 1$, $ \forall k\in\mathcal{K} $. For security enhancement, the BS also injects $J$ columns of unit-power AN signals, denoted by $ \varsigma_j $ with $ j\in \mathcal{J} = \{1, \dots, J\}$. Accordingly, the overall transmit signal vector $\bm{x} \in \mathbb{C}^{M}$ is given by
\begin{equation}
    \bm{x} = \sum\limits_{k \in \mathcal{K}} \bm{u}_k s_k + \sum\limits_{j \in \mathcal{J}} \bm{v}_j \varsigma_j,
\end{equation}
where $\bm{u}_k \in \mathbb{C}^{M}$ is the transmission beamforming vector for user $k$, and $\bm{v}_j \in \mathbb{C}^{M}$ is the $j$-th AN vector. Let $\bm{h}_k \in \mathbb{C}^{M}$ denote the channel gain between the BS and user $k$, and $\bm{h}_{\text{E},k} \in \mathbb{C}^{M}$ as the wiretap channel associated with the corresponding eavesdropper. The signal-to-interference-plus-noise ratio (SINR) for legitimate transmission of user $k$ is
\begin{equation}
    \gamma_k = \frac{
    | \bm{h}_k^{\dagger} \bm{u}_k |^2
    }{
    \sum\limits_{k^\prime \in \mathcal{K}\backslash\{k\}} | \bm{h}_k^{\dagger} \bm{u}_{k^{\prime}} |^2 + 
    \sum\limits_{j\in\mathcal{J}} | \bm{h}_k^{\dagger} \bm{v}_j |^2 + \sigma_0^2
    },
\end{equation}
where $\sigma_0^2$ is the power of the additive white Gaussian noise as assumed identical across all receivers. Similarly, the SINR for the eavesdropper intercepting the signal of user $k$ is
\begin{equation}
    \gamma_{\text{E},k} = \frac{
    | \bm{h}_{\text{E},k}^{\dagger} \bm{u}_k |^2
    }{
    \sum\limits_{k^\prime \in \mathcal{K}\backslash\{k\}} | \bm{h}_{\text{E},k}^{\dagger} \bm{u}_{k^{\prime}} |^2 + 
    \sum\limits_{j\in\mathcal{J}} | \bm{h}_{\text{E},k}^{\dagger} \bm{v}_j |^2 + \sigma_0^2
    }.
\end{equation}
Then, the secrecy rate for user $k$ is obtained as
\begin{equation}
    R_k = \left( \log(1 + \gamma_k) - \log(1 + \gamma_{\text{E},k}) \right)^+,
\end{equation}
where we assume unit bandwidth for the system transmissions and $ (\cdot)^+ = \max\left\{0, \cdot\right\} $.

\begin{figure}[t]
   \centering
   \includegraphics[width=0.40\textwidth]{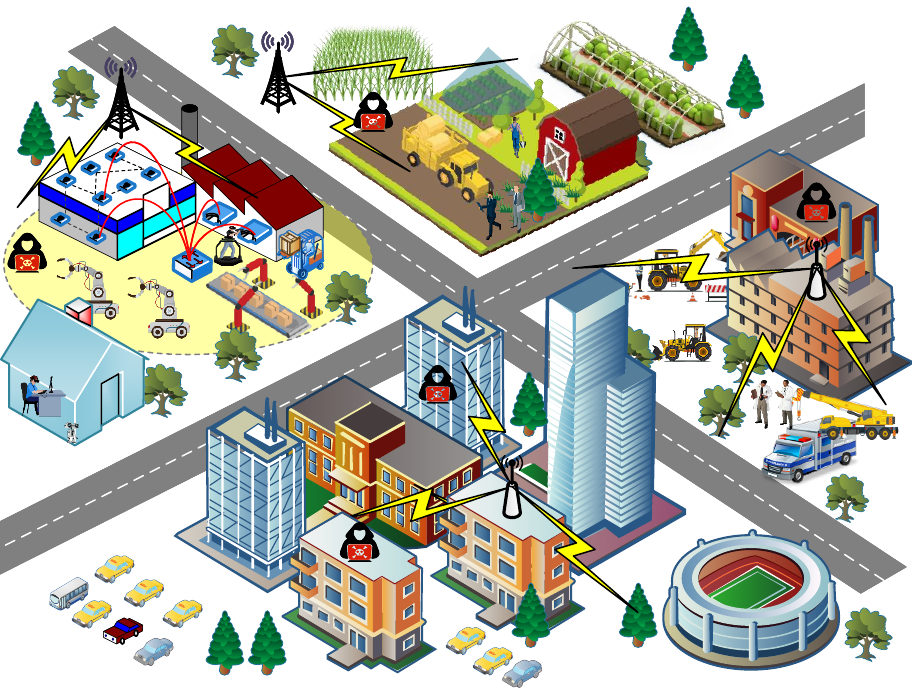} 
   \caption{Secure communications under diverse scenarios.}
   \label{fig:sys}
\end{figure}

\subsection{Problem Formulation}
Our objective is to maximize the sum secrecy rate of all users under a total transmit power constraint. Accordingly, the problem is formulated as
\begin{IEEEeqnarray}{cl} \label{eq:opt}
	\IEEEyesnumber*\IEEEyessubnumber*
	\max_{\{\bm{u_k}\}_{k\in\mathcal{K}}, \{\bm{v_j}\}_{j\in\mathcal{J}}} \quad & \sum_{k \in \mathcal{K}} R_k \label{eq:opt_obj}\\
    \mathrm{s.t.} \quad & \sum\limits_{k \in \mathcal{K}} \left|\bm{u}_k\right|^2 + \sum\limits_{j \in \mathcal{J}} \left|\bm{v}_j\right|^2 \le P_{\max},
\end{IEEEeqnarray}
where $ P_{\max} $ is the power budget of the BS. For notation simplicity, we define a composite vector $ \bm{w} $ that concatenates all transmission and AN beamforming vectors as
\begin{equation}
    \bm{w} = \left[ \bm{u}_1^{\top}, \ldots, \bm{u}_K^{\top}, \bm{v}_1^{\top}, \ldots, \bm{v}_J^{\top} \right]^{\top}, 
\end{equation}
representing the complete security strategy. Also, we define a composite channel vector $\bm{h}$ that collects all channel state information over the network
\begin{equation}
    \bm{h} = \left[ \bm{h}_1^{\top}, \ldots, \bm{h}_K^{\top}, \bm{h}_{\text{E},1}^{\top}, \ldots, \bm{h}_{\text{E},K}^{\top} \right]^{\top}. 
\end{equation}

For the security optimization in~\eqref{eq:opt}, it essentially corresponds to a mapping from $ \bm{h} $ to $\bm{w}$, denoted as $ \bm{w} = \mathcal{F}_{\phi}(\bm{h}) $, such that the overall secrecy rate can be maximized for a given scenario $ \phi $. Although this problem has been well addressed in existing work from both optimization and learning perspectives, they are established upon the implicit assumption of a certain given network scenario. In contrast, we in this work intend to develop a unified security solution to the problem across different scenarios, as shown in Fig.~\ref{fig:sys}.

Towards this goal, we first explicitly define the scenario, which is usually referred to as a high-level concept without quantitative characterization. Considering that different network scenarios induce different propagation, which is further revealed in the channel gains, we use these gains to define the scenario quantitatively. Accordingly, we in this work adopt the Rician channel model specified as
\begin{equation} \label{eq:ch}
    \sqrt{\beta_0 d^{-\kappa_L}} \left( \sqrt{\frac{\kappa_R}{\kappa_R+1}}\bm{h}^{(\text{LoS})} + \sqrt{\frac{1}{\kappa_R+1}}\bm{h}^{(\text{NLoS})} \right),
\end{equation}
where $ \beta_0 $ is the path loss at the reference distance, $ d $ is the distance between the transmitter and receiver, $ \kappa_L $ is the path loss exponent, and $\kappa_R$ is the Rician factor determining the ratio of power in the line-of-sight (LoS) component $\bm{h}^{(\text{LoS})}$ to the non-line-of-sight (NLoS) component $\bm{h}^{(\text{NLoS})}$. The channel model in~\eqref{eq:ch} is versatile to cover a wide variety of wireless environments. On this basis, we define the combination $ \phi = (\kappa_L, \kappa_R) $ as the numerical characteristics to describe the network scenarios. Accordingly, a few typical examples may include that, a tuple (4,1) largely corresponds to the dense urban environment, a tuple (3,1) is likely to be a NLoS suburban scenario, a tuple (2,10) often represents LoS rural or open filed communications, etc.

Under this formulation, the composite channel vector $\bm{h}$ is drawn from a distribution conditioned on the scenario, i.e., $\bm{h} \sim p(\bm{h} | \phi)$. The fundamental challenge under our formulation is no longer to target at a scenario-specific mapping $\mathcal{F}_{\phi}$, but to design a universal framework $\mathcal{F}$ that is effective for a whole set of scenarios $\phi \in \Phi$. The goal is to ensure that for any channel realization $\bm{h}$ across the wireless scenarios in $ \Phi $, the security strategy $\bm{w} = \mathcal{F}(\bm{h})$ achieves high secrecy performance, transforming the problem from scenario-specific optimization to robust cross-scenario generalization.

\section{GDM-Based Security Expert} \label{sec:gdm}

As the basis to unlock the cross-scenario physical layer security framework over $ \Phi $, we first consider a few representative scenarios and establish the scenario-specific security solution, i.e., construct the mapping from the channel condition to the security beamformer as $ \bm{w} = \mathcal{F}_{\phi}(\bm{h}; \bm{\Theta}) $, with given $ \phi\in\Phi $ and $ \bm{\Theta} $ collecting the network parameters. The target output is a high-dimensional continuous security strategy, for which the underlying secrecy-rate maximization is highly non-convex. The non-convexity and random fading lead to diverse sub-optima, motivating distribution learning rather than point estimation. Compared with alternative generative paradigms, conditional diffusion offers stable likelihood-based training, strong capability in modeling complex continuous distributions, and a controllable sampling procedure. Consequently, we train a specialized GDM-based security expert for each scenario, and the expert learns the conditional distribution and generate high-quality security strategies for channel realizations under the current scenario.

\subsection{Conditional Diffusion Model}

\begin{figure}[t]
   \centering
   \includegraphics[width=0.42\textwidth]{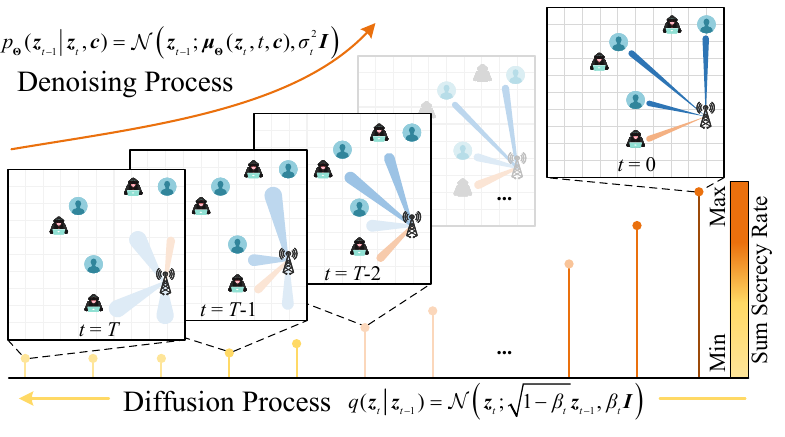} 
   \caption{Conditional denoising diffusion model for security beamforming.}
   \label{fig:gdm}
\end{figure}

The security expert is essentially a conditional denoising diffusion model, which involves two stages, i.e., the forward process that gradually adds noise to the data, and reverse process that remove the noise to generate new data, as shown in Fig.~\ref{fig:gdm}. To facilitate the real-valued neural network operations, we first represent the complex-valued channel and beamforming vectors by separating the real and imaginary parts for new representations as
\begin{equation}
    \bm{z} = \left[
    \Re(\bm{w})^{\top},\ 
    \Im(\bm{w})^{\top}
    \right]^{\top},
    \quad
    \bm{c} = \left[
    \Re(\bm{h})^{\top},\ 
    \Im(\bm{h})^{\top}
    \right]^{\top}.
\end{equation}
Accordingly, the GDM is trained to use $ \bm{c} $ as the condition or guidance to recover the intended security strategy as $ \bm{z} $.

We adopt the channel-conditioned denoising diffusion probability model (DDPM) as the backbone for the GDM expert, which includes $ T $ denoising steps, denoted as $ \mathcal{T} =\{1,2,\cdots, T\} $. Specifically, the forward diffusion process is a Markov chain that progressively adds Gaussian noise to the security strategy tensor over $T$ steps. We use a variance schedule $ \{\beta_t\}_{t\in\mathcal{T}} $ with $ 0\le \beta_t \le 1 $, and conduct the transition at step $t$ as
\begin{equation} \label{eq:fw-t}
    q(\bm{z}_t | \bm{z}_{t-1}) = \mathcal{N}\left(\bm{z}_t; \sqrt{1-\beta_t}\bm{z}_{t-1}, \beta_t\bm{I}\right),
\end{equation}
where $ \bm{I} $ is the identity matrix, and $\bm{z}_0$ is the (near-)optimal security strategy at the initial step. By cascaded iterations in~\eqref{eq:fw-t}, we arrive at a close-form sample $\bm{z}_t$ at any arbitrary step $t$ form from the initialization as
\begin{equation}
    \bm{z}_t = \sqrt{\bar{\alpha}_t}\bm{z}_0 + \sqrt{1-\bar{\alpha}_t}\bm{\epsilon},
\end{equation}
where $\alpha_t = 1-\beta_t$, $\bar{\alpha}_t = \prod_{t^\prime=1}^t \alpha_{t^\prime}$, and $\bm{\epsilon} \sim \mathcal{N}(0, \bm{I})$ is the sampled noise. As the final step $T$ is approached, the distribution of $\bm{z}_T$ reduces to a standard isotropic Gaussian distribution.

The reverse process aims to reconstruct the (near-)optimal secure beamformer from noise. Starting from pure noise $\bm{z}_T \sim \mathcal{N}(0, \bm{I})$, a neural network is trained to predict the noise component $\bm{\epsilon}$ that was added at each step $t\in\mathcal{T}$. Denote the neural network as $\bm{\epsilon}_{\bm{\Theta}}(\bm{z}_t, t, \bm{c})$, it takes the noisy tensor $\bm{z}_t$, the step $t$, and the channel condition $\bm{c}$ as input, and the output is used to assist the evaluation of expectation of denoising step. Specifically, the reverse transition is then given by
\begin{equation}
    p_{\bm{\Theta}}(\bm{z}_{t-1} | \bm{z}_t, \bm{c}) = \mathcal{N}(\bm{z}_{t-1}; \bm{\mu}_{\bm{\Theta}}(\bm{z}_t, t, \bm{c}), \sigma_t^2\bm{I}),
\end{equation}
where the mean $\bm{\mu}_{\bm{\Theta}}$ is a function of the predicted noise $\bm{\epsilon}_{\bm{\Theta}}$ from the neural network
\begin{equation}
    \bm{\mu}_{\bm{\Theta}}(\bm{z}_t, t, \bm{c}) = \frac{1}{\sqrt{\alpha_t}} \left( \bm{z}_t - \frac{1-\alpha_t}{\sqrt{1-\bar{\alpha}_t}} \bm{\epsilon}_{\bm{\Theta}}(\bm{z}_t, t, \bm{c}) \right).
\end{equation}
By iteratively applying this procedure for $T$ steps, the model starts from a pure noise ${\bm{z}}_T \sim\mathcal{N}(\bm{0},\bm{I})$, and arrive the security strategy tensor ${\bm{z}}_0$. Then, we can recover the (near-)optimal beamformer by re-assembling the real and imaginary parts as
\begin{equation}
\begin{aligned}
	\bm{w}^\star = \frac{\sqrt{P_{\max}}}{\left\| \bm{z}_0 \right\|} \left( \bm{z}_0[1: M(K+J)] \right. \qquad\qquad\qquad \\ 
	\left. + \jmath \bm{z}_0[M(K+J)+1:2M(K+J)]  \right),
\end{aligned}
\end{equation}
as normalized by the transmit power budget, where $ \jmath $ is the imaginary unit.

\begin{figure*}[t]
   \centering
   \includegraphics[width=0.85\textwidth]{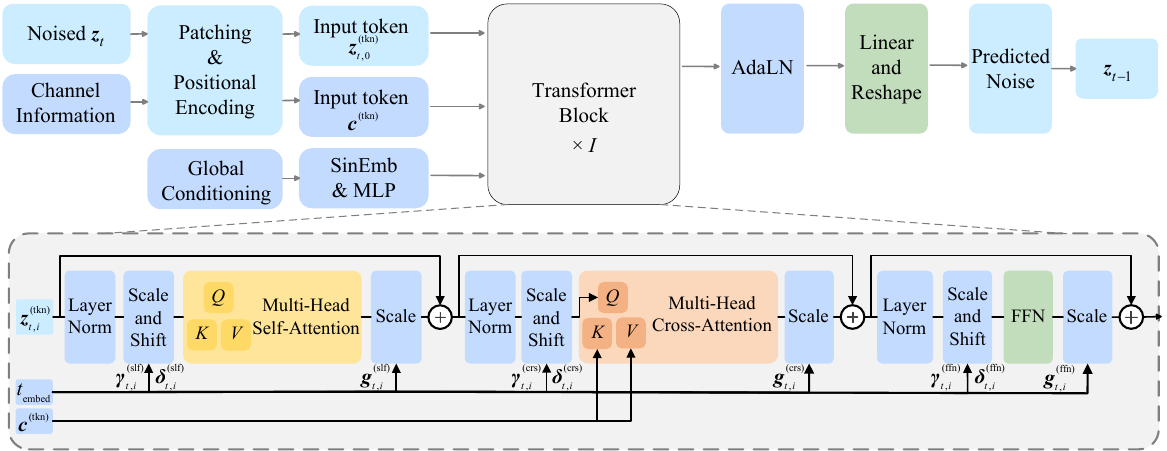} 
   \caption{Transformer-based denoising step.}
   \label{fig:dit}
\end{figure*}

\subsection{Transformer-Based Denoising Process}

The diffusion model presented above consists of $ T $ timestps, where each step requires a neural network to assist the denoising operation. 
In this respect, we adopt the Transformer-based structure, also known as diffusion Transformer (DiT), which is well-suited to the considered security problem, capturing the intricate influence among different users while conditioned on the network channel state. Moreover, the DiT has the advantages of permutation-invariant attention mechanism to handle variable-length input, and thus improves the generalization capability of the model.

Particularly, this security-aware DiT composes of $I$ cascaded Transformer blocks, denoted as $ \mathcal{I} = \{1,2,\ldots,I\} $, shown in Fig.~\ref{fig:dit} and elaborated below.

\subsubsection{Input Embedding}
For Transformer-based operations, the first step is to tokenize the input.

Consider the $t$-th denoising step, the input noisy strategy tensor $\bm{z}_t$ and the channel condition tensor $\bm{c}$ are treated as 2D grids. We use a patching strategy where each entry (corresponding to a specific antenna and beamforming vector) is linearly projected into a latent token, with a trainable convolutional layer. The token is further flattened and added with a sinusoidal positional encoding to retain the spatial information, leading to the tokenized input as $ \bm{z}_{t,0}^{(\text{tkn})} $ and $\bm{c}^{(\text{tkn})}$. Accordingly, the tokenization is specified as
\begin{equation}
	\bm{z}_{t,0}^{(\text{tkn})} = \mathsf{FLT} \left(\mathsf{Conv2D}\left( \bm{z}_t  \right)\right) + \bm{e}^{(\text{pos})},
\end{equation}
and
\begin{equation}
	\bm{c}^{(\text{tkn})} = \mathsf{FLT} \left(\mathsf{Conv2D}\left( \bm{c}  \right)\right) + \bm{e}^{(\text{pos})},
\end{equation}
where the subscript 0 indicates the initial input for the cascaded $I$ blocks, $\mathsf{FLT}(\cdot)$ denotes the flattening operation, and $ \bm{e}^{(\text{pos})} $ is the positional encoding shared at the noisy input and channel condition. The tokenization explicitly preserves the security strategy structure, which is critical for scaling the expert to different network configurations. Meanwhile, the positional embedding is designed to align token positions with user-antenna streams, providing a physically meaningful inductive bias that improves model generalization capability.

Moreover, the information on current step, number of users, and number of antennas, are embedded as the global conditioning vector. In this regard, the denoising operation becomes well aware of current step and the basic network configuration, such that better generalization can be supported. Particularly, the step $t$ goes through the sinusoidal embedding, together with the scalar $M$ and $K$, constitute a vector as the global conditioning embedding. Furthermore, a linear projection layer is employed to create a tuple of representations used later in the Transformer blocks, specified as
\begin{equation}
\begin{aligned}
	\left\{ \bm{\gamma}^{(\text{slf})}_{t,i}, \bm{\delta}^{(\text{slf})}_{t,i}, g^{(\text{slf})}_{t,i}, \bm{\gamma}^{(\text{crs})}_{t,i}, \bm{\delta}^{(\text{crs})}_{t,i}, g^{(\text{crs})}_{t,i},\bm{\gamma}^{(\text{ffn})}_{t,i}, \bm{\delta}^{(\text{ffn})}_{t,i}, g^{(\text{ffn})}_{t,i}  \right\}_{i\in\mathcal{I}} \\ = \mathsf{LIN}([\mathsf{SinEmb}(t), M, K]),
\end{aligned}
\end{equation}
where $ \bm{\gamma}^{(\text{x})}_{t,i} $ and $ \bm{\delta}^{(\text{x})}_{t,i} $ are the scale and shift for the adaptive layer normalization (AdaLN) operations, and $ g^{(\text{x})}_{t,i} $ indicates the gating coefficient, with $ \text{x} \in \{\text{slf}, \text{crs}, \text{ffn}\} $ denotes the self-attention, cross-attention, and feed forward network within the Transformer. As such, the information on current denoising step and global network setting can be conveyed into the denoising process, improves both fitness and generalization of the model.

\subsubsection{DiT Block Structure}
As noted before, the tokenized input goes through $I$ DiT blocks, which mainly resort to the attention mechanism to integrating the channel state information to refine the security strategy. The main operations include:
\begin{itemize}
	\item \textbf{AdaLN}: As the basic supporting operation within the Transformer, we adopt the adaptive LayerNorm, which exploits the global conditioning embedding induced coefficients. Particularly, for the input tensor $\bm{y}$, adaptive LayerNorm is conducted as
	\begin{equation}
	\begin{aligned}
		\mathsf{AdaLN}\left(\bm{y}; \bm{\gamma}^{(\text{x})}_{t,i}, \bm{\delta}^{(\text{x})}_{t,i}\right) = \bm{\gamma}^{(\text{x})}_{t,i} \odot \mathsf{LN}(\bm{y}) + \bm{\delta}^{(\text{x})}_{t,i}, \quad\\ \text{x} \in \{\text{slf}, \text{crs}, \text{ffn}\},
	\end{aligned}
	\end{equation}
	as will be used later in the DiT, where $ \mathsf{LN}(\cdot) $ is the LayerNorm operation.

	\item \textbf{Self-attention}: As the tokenized input is modulated by the AdaLN, self attention calculation is conducted. This mechanism allows the model to capture the complex-coupled and long-range dependencies among the strategy elements of the beamforming and AN vectors. Then, the attention results are updated through a gated residual to update the token, given as
    \begin{equation}
        \bm{z}_{t,i}^{(\text{tkn})} \gets \bm{z}_{t,i}^{(\text{tkn})} + g^{(\text{slf})}_{t,i} \mathsf{SlfAtt}(\mathsf{AdaLN}(\bm{z}_{t,i}^{(\text{tkn})}; \bm{\gamma}^{(\text{slf})}_{t,i}, \bm{\delta}^{(\text{slf})}_{t,i})),
    \end{equation}
    where $ \mathsf{SlfAtt}(\cdot) $ denotes the self-attention calculation and the query, key, and value sequences are projections of input tokens.

    \item \textbf{Cross-attention}: Then, the tokenized channel state information is exploited as the guidance for the denoising process. Particularly, the cross-attention operation is conducted that, the output of self-attention serves as the query sequence, and the keys and values are based on the channel tokens, specified as
    \begin{equation}
    \begin{aligned}
        &\bm{z}_{t,i}^{(\text{tkn})} \gets \bm{z}_{t,i}^{(\text{tkn})} + g^{(\text{crs})}_{t,i} \\
        &\qquad\times\mathsf{CrsAtt}(\mathsf{AdaLN}(\bm{z}_{t,i}^{(\text{tkn})}; \bm{\gamma}^{(\text{crs})}_{t,i}, \bm{\delta}^{(\text{crs})}_{t,i}), \bm{c}^{(\text{tkn})}),
    \end{aligned}
    \end{equation}
    where $ \mathsf{CrsAtt}(\cdot) $ denotes the cross-attention. Injecting channel information through conditional tokens and cross-attention enable the expert to learn a channel-aware solution landscape, which is necessary for robust secrecy strategy generation under heterogeneous propagation conditions.

    \item \textbf{Feed-forward}: Finally, the tokens pass through a position-wise feed-forward network, which consists of two linear layers. The gated residual is also applied for the output, represented as
    \begin{equation}
        \bm{z}_{t,i}^{(\text{tkn})} \gets \bm{z}_{t,i}^{(\text{tkn})} + g^{(\text{ffn})}_{t,i} \mathsf{FFN}(\mathsf{AdaLN}(\bm{z}_{t,i}^{(\text{tkn})}; \bm{\gamma}^{(\text{ffn})}_{t,i}, \bm{\delta}^{(\text{ffn})}_{t,i})),
    \end{equation}
    where $ \mathsf{FFN}(\cdot) $ denotes the feed forward operation.
\end{itemize}
The DiT operations above are conducted $ I $ times within the $ I $ cascaded blocks, the channel condition is repeatedly injected to affect the strategy token such that the security strategy get continuously adjusted and refined.

\subsubsection{Output Generation}
After passing through all $I$ DiT blocks, the final sequence of strategy tokens, i.e., $\bm{z}_{t,I}^{(\text{tkn})}$, is arrived. We then reconstruct the updated noisy strategy vector from the arrived token. To this end, we also adopt the AdaLN operation to modulate the token, which is then linearly projected to produce the predicted noise with the same dimension of the strategy vector. This finishes the overall noise prediction network $\bm{\epsilon}_{\bm{\Theta}}$. The noise prediction is further used by the sampling algorithm to compute the denoised mean $ \bm{\mu}_{\bm{\Theta}} $, leading to a less noisy strategy vector $ \bm{z}_{t-1} $, as the output of current denoising step.

\subsection{Loss Function and Training}

The training of each GDM-based security expert follows a two-stage paradigm designed to combine the stability of supervised learning with the performance-oriented fine-tune.

In the first stage, we perform supervised pre-training. For each representative scenario $\phi$, we adopt the security optimization algorithm (e.g., the approach in~\cite{dat}) to construct a dataset of channel realizations and their corresponding near-optimal security strategies. The GDM expert is then trained to learn the conditional distribution $p(\bm{w}|\bm{h})$ by minimizing the standard diffusion model loss, as the mean squared error between the true and predicted noise, specified as
\begin{equation}
    \mathcal{L}^{(\text{pre})}\left(\bm{\Theta}\right) = \mathbb{E}_{t, \bm{w}, \bm{\epsilon}} \left[ \|\bm{\epsilon} - \bm{\epsilon}_{\bm{\Theta}}(\bm{z}_t, t, \bm{c})\|^2 \right].
\end{equation}
This pre-training stage enables the model with a fundamental understanding of the solution space.

In the second stage, we fine-tune the pre-trained model to further maximize the sum secrecy rate. Particularly, for a given channel condition, we use the GDM expert to generate a security strategy. We use the negative of sum secrecy rate as the loss function, given as
\begin{equation}
    \mathcal{L}^{(\text{ftn})}(\bm{\Theta}) = \mathbb{E}_{\bm{h}}\left\{ - \sum_{k \in \mathcal{K}} R_k(\bm{h}, \mathcal{F}_{\phi}(\bm{h}; \bm{\Theta}))\right\},
\end{equation}
such that the model parameters are then adjusted to further improve the secrecy rate.

\subsection{Complexity of GDM Experts}

For the proposed GDM experts, each performs DDPM sampling with $T$ denoising steps, where each step evaluates a DiT denoiser consisting of $I$ cascaded Transformer blocks. The DiT block includes the operations of AdaLN, self-attention over the strategy tokens, cross-attention between the strategy and channel tokens, and position-wise feed-forwards. For the tokenization operations, the length of strategy token and channel token are $L^{\text{(z)}}=M(K+J)$ and $L^{\text{(c)}}=2MK$, respectively. Also, denote the token embedding dimension and the feed-forward hidden width by as $d^{\text{(tkn)}}$ and $d^{\text{(ffn)}}$, respectively, the module-specific complexity is analyzed as follows. First, the self-attention over $L^{\text{(z)}}$ strategy tokens has complexity $ \mathcal{O}\left( {(L^\text{(z)})}^2d^{\text{(tkn)}} + {L^\text{(z)}}{(d^{\text{(tkn)}})}^2 \right) $, due to the attention weights and output projection calculation. Also, the cross-attention is of complexity $ \mathcal{O} \left(L^{\text{(z)}}L^{\text{(c)}}d^{\text{(tkn)}} + \left(L^{\text{(z)}}+L^{\text{(c)}}\right)(d^{\text{(tkn)}})^2\right) $, due to the calculation between strategy tokens and channel tokens. Finally, the feed-forward network has complexity $ \mathcal{O} \left(L^{\text{(z)}} d^{\text{(tkn)}} d^{\text{(ffn)}}\right) $. Therefore, the complexity of each GDM expert is given as
\begin{equation}
\begin{aligned}
\mathcal{O}\left(TI\left( (L^\text{(z)})^2d^{\text{(tkn)}} + L^{\text{(z)}}L^{\text{(c)}}d^{\text{(tkn)}} \qquad\qquad\qquad\:\: \right.\right. \\ \left.\left.  + (2L^{\text{(z)}}+L^{\text{(c)}})(d^{\text{(tkn)}})^2 + L^{\text{(z)}} d^{\text{(tkn)}} d^{\text{(ffn)}} \right)
\right),
\end{aligned}
\end{equation}
which is further scaled with the number of selected experts when assembled within the overall MoE architecture.

\section{MoE-Based Cross-Scenario Security Solution} \label{sec:moe}

In the preceding section, we elaborated on the design of a scenario-specific GDM-based security expert. Although a single Transformer-based diffusion model possesses a certain degree of scalability, its security performance degrades significantly when the underlying wireless environment changes. To effectively provide security across a continuum of scenarios in $\Phi$, we propose a MoE framework that leverages the pre-trained GDM-based models as its supporting experts.

\begin{figure*}[t]
   \centering
   \includegraphics[width=0.65\textwidth]{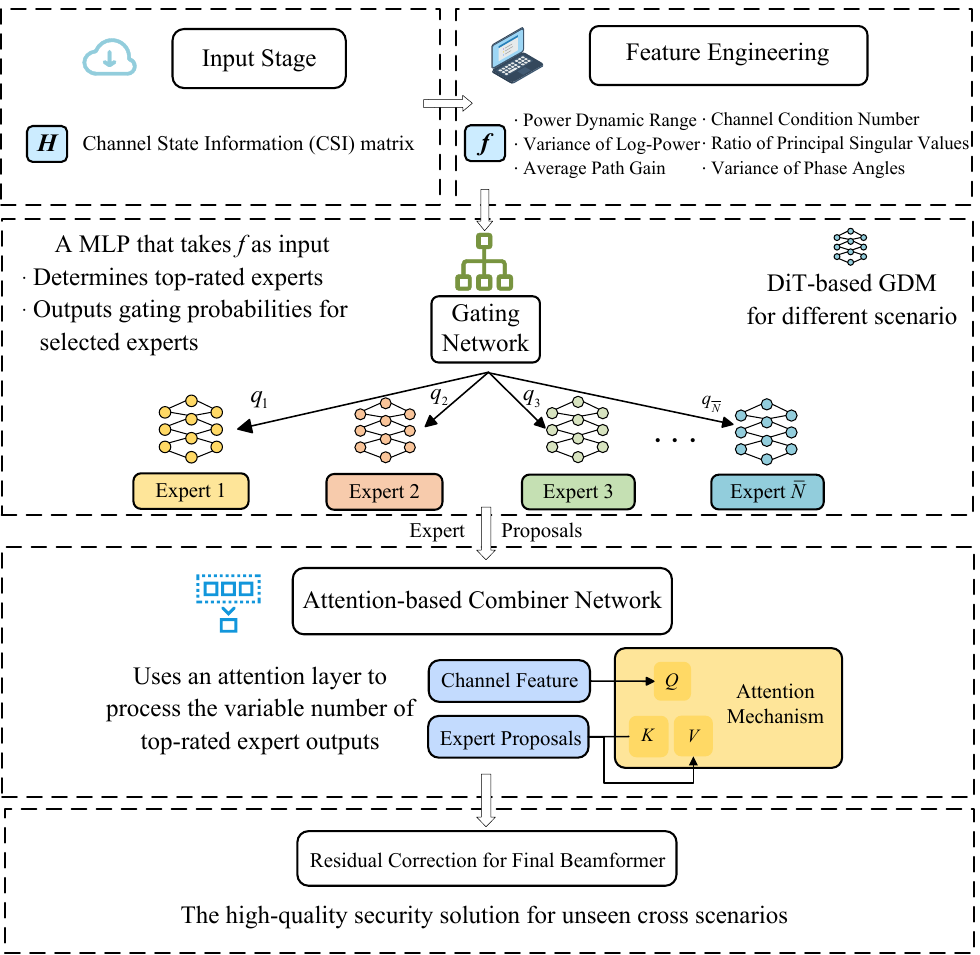} 
   \caption{The MoE framework with GDM security experts.}
   \label{fig:moe}
\end{figure*}

To this end, we first select a set of representative wireless scenarios, indexed by $\mathcal{N} = \{1,2,\ldots,N\} $, with selected scenario denoted by $\Phi^{(\text{rep})} = \{\phi_n\}_{n\in\mathcal{N}}$. For each specific scenario $\phi_n \in \Phi^{(\text{rep})}$, we construct and train a dedicated GDM-based expert, possessing the learned security strategies, denoted by $\mathcal{F}_{\phi_n}$. The MoE framework is then established by orchestrating this collection of experts, in order to combine the specialized knowledge of these experts to achieve robust and high-performance security provisioning across the entire space of scenarios $\Phi$. 

Therefore, a gating network, or router is established as the core of the MoE framework, which intends to accurately identify the input channels information with its scenario properties. Then, the corresponding experts can be called such that their knowledge can be synergized to produce the scenario-aware security solution. The overall design is shown in Fig.~\ref{fig:moe} and the details are elaborated as follows.

\subsection{Scenario-Aware Feature Engineering}

The responsibility of the router is to distinguish current network scenario and call the right experts. However, the scenario properties are embedded in the network channel information, which is of high-dimension and inherently random. Since our considered scenario are mainly characterized by the poth loss exponent and Rician factor, we introduce the feature engineering that are of low dimension yet remain sensitive to the physical parameters. Particularly, we construct a 6-dim feature, denoted by $\bm{f},$ by calculating the statistical characteristics of the channel information.

The first category includes 3 features capturing the power distribution, which are sensitive to the path loss-related component, specified as
\begin{itemize}
 	\item \textbf{Power Dynamic Range} ($f_1$): The ratio of the maximum to minimum channel gain among the receivers, i.e., 
 	$\max\limits_{k\in\mathcal{K}}\left\{\|\bm{h}_k\|^2,\|\bm{h}_{\text{E},k}\|^2\right\} / \min\limits_{k\in\mathcal{K}}\left\{\|\bm{h}_k\|^2,\|\bm{h}_{\text{E},k}\|^2\right\}$. A higher ratio leads to a more extreme power range.
    \item \textbf{Variance of Log-Power} ($f_2$): The variance of the log-scaled channel gain, $\mathsf{Var}\left(\log\left(\left[\|\bm{h}_k\|^2,\|\bm{h}_{\text{E},k}\|^2\right]_{k\in\mathcal{K}}\right)\right)$, which quantifies the power disparity across users.
    \item \textbf{Average Path Gain} ($f_3$): The mean received power across all users, $\mathsf{Mean}\left(\left[\|\bm{h}_k\|^2,\|\bm{h}_{\text{E},k}\|^2\right]_{k\in\mathcal{K}}\right)$, reflecting the overall channel power attenuation.
\end{itemize}

The second category has 3 features probing into the spatial structure of the channel condition, which is highly concerned with the LoS component of the channel, specified as
\begin{itemize}
	\item \textbf{Channel Condition Number} ($f_4$): We concatenate the channel vectors at the users as a matrix, $\bm{H} = \left[\bm{h}_k, \bm{h}_{\text{E},k}\right]_{k\in\mathcal{K}}$, and calculate the ratio of the largest to the smallest singular value of $\bm{H}$, i.e., $\delta_{\max}(\bm{H}) / \delta_{\min}(\bm{H})$, where $ \delta(\cdot) $ denotes the singular value. A higher Rician factor often leads to a more ill-conditioned, lower-rank channel matrix.
    \item \textbf{Ratio of Principal Singular Values} ($f_5$): The ratio of the largest to the second largest singular value, $\delta_1(\bm{H}) / \delta_2(\bm{H})$, which more directly measures the dominance of the strongest spatial path.
    \item \textbf{Variance of Phase Angles} ($f_6$): The average phase variance across the antennas for each user's channel vector, i.e., $\mathsf{Mean}(\mathsf{Var}(\mathsf{Arg}(\bm{H})))$, where $ \mathsf{Arg}(\cdot) $ denotes the element-wise angle calculation and $ \mathsf{Var}(\cdot) $ is conducted over each column. In this respect, a stronger LoS component induces higher phase coherence and thus lower overall variance.
\end{itemize}

Through the engineered feature $ \bm{f} $, we obtain a compact yet comprehensive representation of the channel characteristics, which can be then used for scenario classification and expert selection. Moreover, the feature engineering enables a fix-dimension representation regardless of the network channel dimensions, making the router directly applicable for networks with variable size configurations.

\subsection{Router Design and Training}

The router, intending for scenario identification, can be implemented as a standard MLP for classification task. Accordingly, the router takes the channel feature vector $ \bm{f} $ as input, and outputs a probability distribution over the $N$ classes of representative scenarios.

The training of the router follows a standard supervised manner. To this end, we first generate a labeled dataset. Specifically, for each representative scenario $ \phi_n\in\Phi^{(\text{rep})} $, we generate a set of channel condition matrices following the scenario setting, i.e., generate the random network topology and calculate the channel gain based on the path loss exponent and Rician factor determined by the scenario. Then, we calculate the engineered channel feature, together with the ground-truth label associated with the representative scenario, is used for router training. The training is conducted to minimize the cross-entropy loss. The final layer of the router consists of $N$-dim output with a $\mathsf{Softmax}$ activation function, which produces a gating probability $N$-dim vector $ \bm{q} = [q_n]_{n\in\mathcal{N}} $ such that $ \sum_{n\in\mathcal{N}}q_n = 1 $, and $ q_n $ is the probability that scenario $ \phi_n $ is identified and expert $ \mathcal{F}_{\phi_n}  $ gets selected.

\subsection{Expert Selection and Combination}

As the router produce a probability distribution over the experts, the final MoE operation is to select the experts properly and determine the security beamforming strategy. This can be generally categorized into two cases. First, for scenarios that closely align with the representative ones, we can safely expect that one element of the gating probability vector is close to 1, which thus leads to a clear choice. In this case, we can simply select the corresponding expert for security strategy determination.

For the second case, also more likely to be encountered, there are unseen cross-scenarios, for which the gating probability also indicates multiple relevant experts. In this case, we adopt the ``top-\textit{k}'' gating principle, where a set of $ \bar{N} $ top-rated experts are selected, denoted as $ \bar{\mathcal{N}} \subseteq \mathcal{N} $. A direct approach is to normalize the gating probability within $ \bar{\mathcal{N}} $ and use it as the weights to reach a linear combination, specified as
\begin{equation} \label{eq:w-star}
	\bm{w}^{\star} = \sum_{n \in \bar{\mathcal{N}}} \frac{q_n}{\sum_{n^\prime \in \bar{\mathcal{N}}} q_{n^\prime}} \mathcal{F}_{\phi_n}(\bm{h}),
\end{equation}
which is further normalized with respect to the transmit power constraint as
\begin{equation} \label{eq:w-norm}
	\bm{w}^{\star} \gets \bm{w}^{\star} \sqrt{\frac{P_{\max}}{\|\bm{w}^{\star}\|^2}}.
\end{equation}
As a further note, we can see that the first case with one selected expert corresponds to the ``top-one'' rule, as the special case to the general ``top-\textit{k}'' settlement.

\subsection{Attention-Based Combiner Network}

The previous combination approach based on the ``top-\textit{k}'' rule, through direct and efficient, can be inherently limited, as it has the implicit assumption of linear interpolation. Given the complicated relationship between the considered scenario factors, i.e., path loss exponent and Rician factor, and the resultant channel gain, the assumption that the optimal solution lies on the straight line between the experts can be far from reasonable. Consequently, we resort to some more powerful mechanism to learn such non-linear combinations, towards more effective security enhancement.

To address this limitation and for improved security performance, we introduce a dedicated combiner network with attention mechanism as a residual learner. Instead of directly learning the final beamformer, the combiner is trained to predict a residual correction term, which is used along with the weighted expert output for final beamforming determination. In this respect, the combiner adjustment can be regarded as a further fine-tune to learn the non-linear mapping from the weighted experts to the optimum, while enabling a scalable model to handle variable numbers of experts.

\subsubsection{Attention-Based Architecture}

Basically, the number of selected experts through the gating network can be different, depending on how ``cross'' is the considered scenario, as compared with the representative scenarios. With this in mind, the combiner network is expected to synergize the output from different numbers of active experts. In this regard, a standard MLP with fixed-size input is incapable for such cases. We thus resort to the permutation-invariant attention mechanism to tackle the variable-length input.

Specifically, the input of combiner network includes:
\begin{itemize}
	\item The scenario-aware feature vector, $ \bm{f} $.
	\item The beamformer proposal by the selected experts, $\{\bm{w}_n\}_{n \in \bar{\mathcal{N}}}$, along with the selection probability, $ \bm{q} $.
\end{itemize}
Then, the input information is projected into the attention space and reinterpreted as query, key, and value. Here, we use the feature vector to induce the query through a linear network, given as $ \bm{Q} = \mathsf{LIN}^{(\text{qry})}(\bm{f}) $. Meanwhile, we use the weighted beamformer vector from the experts to derive the key and value, given as $ \bm{K}_n = \mathsf{LIN}^{(\text{key})}(\bm{z}_n) $ and $ \bm{V}_n = \mathsf{LIN}^{(\text{val})}(\bm{z}_n) $, where $ \bm{z}_n $ is the beamforming representation in real space as $ \bm{z}_n = q_n\cdot\left[\Re(\bm{w}_n)^{\top},\Im(\bm{w}_n)^{\top}\right]^{\top} $, for all $ n\in\bar{\mathcal{N}} $. Note that the query is unique while the each expert output is projected to an independent key and value.

With the query, key, and value obtained, we can then conduct the attention-based aggregation such that the expert proposals can be synthesized in the form of
\begin{equation} \label{eq:comb-att}
    \bm{o}^{(\text{att})} = \sum_{n \in \bar{\mathcal{N}}} \left( \mathsf{Softmax}\left(\frac{\bm{Q} \bm{K}_n^{\top}}{\sqrt{d^{(\text{att})}}}\right) \bm{V}_n \right),
\end{equation}
where $ d^{(\text{att})} $ is the dimension of the latent attention space. As we see from~\eqref{eq:comb-att}, the attention mechanism allows the combiner to exploit the scenario feature (query) to weight the importance of the expert proposals, and gradually learns the non-linear relationship between the directly weighted beamformer and the optimum.

The context vector $ \bm{o}^{(\text{att})} $ from the attention is then fed into a small MLP to generate the residual correction to the real-valued representation of the weighted beamforming, given as
\begin{equation}
    \Delta\bm{z} = \mathsf{MLP}^{(\text{rdo})}(\bm{o}^{(\text{att})}).
\end{equation}
Here the MLP adopted is of fixed-dimension input and output, and the residual beamformer is reconstructed as $ \Delta\bm{w} = \left[ \Delta\bm{z}[1: M(K+J)] +\jmath \Delta\bm{z}[M(K+J)+1: 2M(K+J)]  \right] $. Then, we can use the weighted expert output along the residual beamformer for the final output, given as
\begin{equation}
    \bm{w}^{\star} = \bm{w}^{\circ} + \Delta\bm{w},
\end{equation}
where $ \bm{w}^{\circ} $ is obtained from the experts in the same form of~\eqref{eq:w-star}, and $ \bm{w}^{\star} $ is similarly normalized as~\eqref{eq:w-norm} to obtain the final beamforming vector.

\subsubsection{Combiner Network Training}

For the overall MoE landscape, the combiner training accounts for the final stage, before which the GDM-based experts and gating network has already been trained with frozen neural network parameters. To fully reveal the potential for superior security performance, we adopt the unsupervised learning which also alleviates the burden of data generation and collection.

Specifically, for the cross-scenario with unseen settings of scenario parameters, we randomly generate the network topology and calculate the channel vector of the network. The channel vector gets through the router to get the probability distribution of the experts. Under given ``top-\textit{k}'' rule, the corresponding top-rated experts are called to output their beamforming proposals. Finally, the channel feature and the expert output are fed into the combiner for the beamformer output. The combiner network is trained to maximize the sum secrecy rate as
\begin{equation}
    \mathcal{L}^{(\text{cmb})}(\bm{\Xi}) = \mathbb{E}_{\bm{h}}\left\{ - \sum_{k \in \mathcal{K}} R_k(\bm{w}; \bm{\Xi}))\right\},
\end{equation}
where $ \bm{w} $ is the combiner output and $ \bm{\Xi} $ collects the parameters of the combiner network.

Overall, the residual learning provides a direct path to approach the optimal secure beamforming. This process trains the combiner to explicitly learn the non-linear correction to transform a simple linear interpolation towards a high-fidelity security strategy. Since the residual learning starts from an already strong baseline, the learning can be conducted in an efficient manner to cover the security requirement for the unseen cross-scenarios.

\subsection{Complexity of the Router and Combiner}

Based on the preceding discussions, the router is a simple MLP operated over the engineered features. To obtain the feature vector, the matrix norm/mean/variance has complexity $ \mathcal{O}(MK) $, while the singular value decomposition is of complexity $ \mathcal{O} \left(\min(M,2K)^2\max(M,2K)\right) $. Then the MLP-based router has complexity $ \mathcal{O}\left(L^{\text{(rtr)}}(d^{\text{(rtr)}})^2\right) $ with $ L^{\text{(rtr)}} $ and $ d^{\text{(rtr)}} $ being the number and dimension of the hidden layers, respectively. Accordingly, the router complexity is $ \mathcal{O} \left(MK + \min(M,2K)^2\max(M,2K) + L^{\text{(rtr)}}(d^{\text{(rtr)}})^2\right) $.

For the attention-based combiner, the attention coefficients are calculated based on the engineered features and proposals from the selected experts, and thus the complexity is $ \mathcal{O}\left( \bar{N}\cdot2M(K+J)d^{\text{(att)}}  \right) $. Then the residual correction calculation through the MLP has complexity $ \mathcal{O}\left(L^{\text{(rdo)}}(d^{\text{(rdo)}})^2\right) $ with $ L^{\text{rdo}} $ layers of $ d^{\text{rdo}} $-dim neurons. Therefore, the combiner complexity is $ \mathcal{O}\left( \bar{N}\cdot2M(K+J)d^{\text{(att)}} +L^{\text{(rdo)}}(d^{\text{(rdo)}})^2 \right) $.

\section{Simulation Results} \label{sec:sim}

Simulation results are provided to evaluate our proposed framework for cross-scenario security performance. We consider a multi-user MISO system within an area of 500~m$\times$500~m, where the BS is at the center and the legitimate users randomly distributed with an average distance of 100~m, and the eavesdroppers are randomly located with an average of 20~m away from its corresponding receiver. The base station has a maximum transmit power of 20~dBm, and the background noise power at the receivers is -110~dBm. The number of legitimate users (and thus eavesdroppers) is 4, and the number of BS antennas is 8. These settings are used as default unless otherwise indicated. For scenario setting, we consider a set of path loss exponents as \{2,3,4\}, and a set of Rician factors as \{1,10\}, which constitute totally 6 representative scenarios with different combination of path loss exponents and Rician factors.

For the neural networks, the proposed framework is trained on a dataset of 48,000 samples, with 8,000 samples generated for each of the 6 representative channel scenarios. Each GDM-based expert has 6 DiT blocks, trained for 2,000 epochs using the Adam optimizer with a learning rate of 5$\times$10\textsuperscript{-5}. The diffusion process uses 100 steps with $\beta_t$ scheduled from 0.0001 to 0.02. The DiT attention modules have 8 heads and a token dimension of 512, while its feed-forward network uses two linear layers of 512$\rightarrow$1536$\rightarrow$512 with GELU activation. The gating network is an MLP with three linear layers of hidden dimension 512, trained for 10 epochs with a learning rate of 5$\times$10\textsuperscript{-4}. The combiner has an attention dimension of 256 and a three-layer MLP with hidden dimension 512, trained for 200 epochs with a learning rate of 5$\times$10\textsuperscript{-5}.

Since the overall MoE framework is complicated, to make the role of each component explicit, the results are organized as a component-wise evaluation pipeline. We first evaluates the GDM experts for the representative scenarios under varying system configurations. Then, the router is validated via feature separability and expert selection accuracy. Finally, the end-to-end MoE performance across representative and unseen cross-scenarios are corroborated considering different combiners with a sensitivity study on different expert activations.

\begin{figure}[t]
   \centering
   \includegraphics[width=0.48\textwidth]{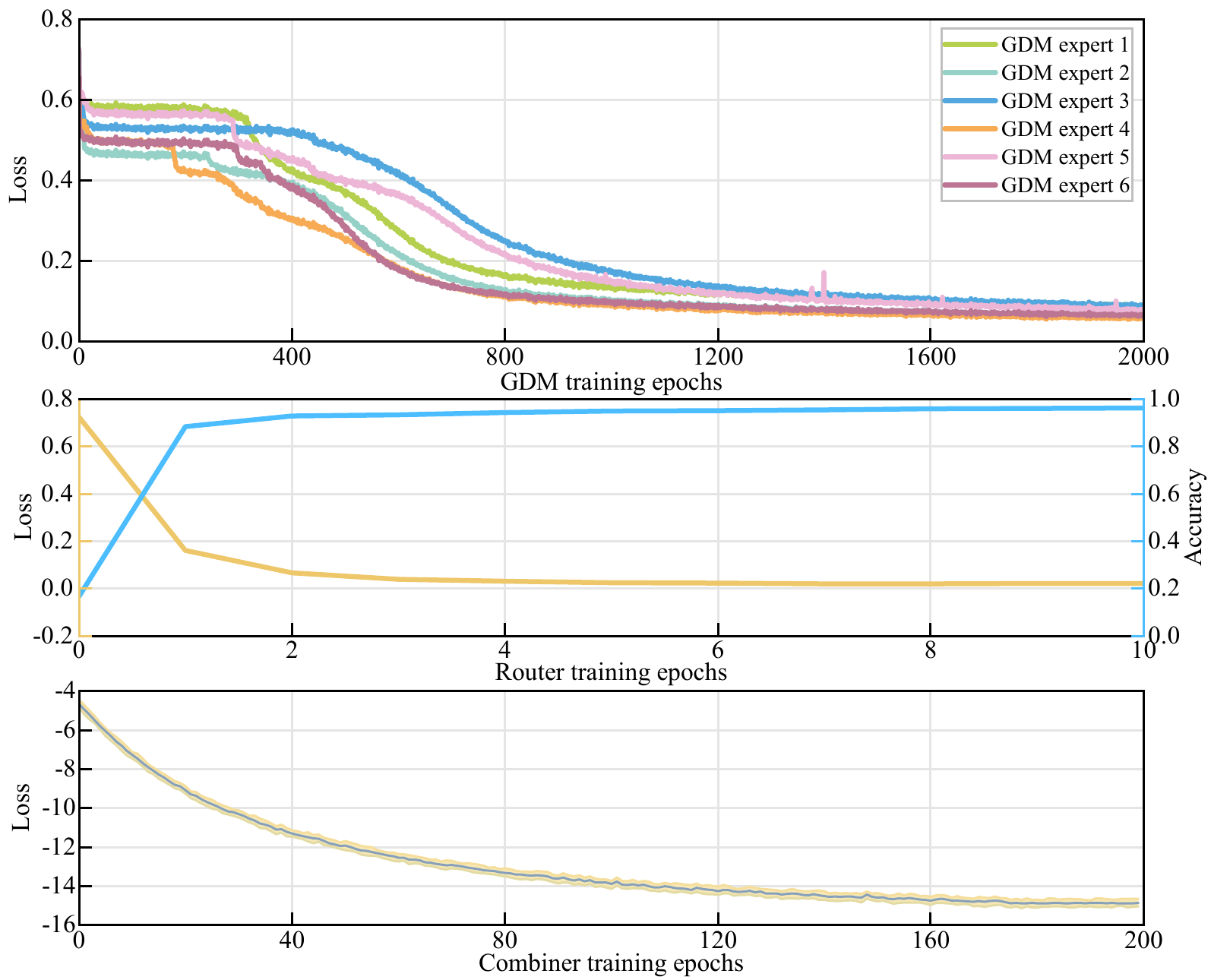} 
   \caption{The training convergence of experts, router, and combiner.}
   \label{fig:loss}
\end{figure}

\subsection{Convergence and Complexity Validation}

We first illustrates the training convergence of the three main components in the proposed MoE framework in Fig.~\ref{fig:loss}. The losses of the six GDM experts decrease steadily with training epochs and converge to low and stable values, indicating reliable learning of scenario-specialized generative security strategies. The router converges within a few epochs, where the loss drops rapidly and the classification accuracy quickly approaches saturation, validating the effectiveness of the proposed low-dimensional feature for scenario recognition. Moreover, the combiner loss decreases smoothly and stabilizes along with the trainings, demonstrating that the attention-based residual fusion can be learned in a stable manner to refine the selected expert proposals. These convergence results collectively suggest that the overall framework is trainable with stable convergence behaviors across all modules.

To quantify the model complexity and inference latency, we evaluate the number of active parameters, computation FLOPs, and inference time, under different simulation settings, and the results are summarized in Table~\ref{tab}. A key insight is that while the total framework has a comprehensive parameter set of 133.03 M, to ensure the coverage across diverse scenarios, the sparse gating mechanism ensures that the computational load remains relatively lightweight. Specifically, under Top-1 expert selection rule, the system activates only 23.23 M parameters, achieving an ultra-low inference latency of a few milliseconds. Even when scaling to Top-4 strategy, the latency remains millisecond level. Moreover, for a fixed number of selected experts, changing the network size leads to a moderate increase in FLOPs and latency due to the enlarged strategy and channel token lengths. Overall, the results indicate that, while the framework contains multiple experts, its online complexity is tunable and remains in the millisecond range for the considered settings, supporting the practicality of the proposed cross-scenario design.

\begin{table}[t]
\caption{Inference Complexity and Latency Profile under Different Network Configurations}
\label{tab}
\centering
\begin{tabular}{c c c c c}
\toprule
Top-\textit{k} & Netw. Config. & Act. Para. & FLOPs & Infer. Tm. (ms) \\
\midrule
      & $M$ = 8, $K$ = 2  &         & 1.00 G & 5.926 \\
Top-1 & $M$ = 8, $K$ = 4  & 23.23 M & 1.72 G & 6.424 \\
      & $M$ = 6, $K$ = 4  &         & 1.29 G & 6.043 \\
\midrule
      & $M$ = 8, $K$ = 2  &         & 2.02 G & 11.129 \\
Top-2 & $M$ = 8, $K$ = 4  & 45.19 M & 3.45 G & 12.126 \\
      & $M$ = 6, $K$ = 4  &         & 2.59 G & 11.364 \\
\midrule
      & $M$ = 8, $K$ = 2  &         & 4.02 G & 21.536 \\
Top-4 & $M$ = 8, $K$ = 4  & 89.11 M & 6.90 G & 23.532 \\
      & $M$ = 6, $K$ = 4  &         & 5.15 G & 22.006 \\
\midrule
\multicolumn{5}{l}{
  \parbox{8cm}{\raggedleft
Model size: 0.14 M (Router) + 1.13 M (Combiner) + \\6$\times$21.96 M (GDM); Total 133.03 M
}} \\
\bottomrule
\end{tabular}
\end{table}

\begin{figure*}[t] 
  \centerline{
  \subfigure[Performance with respect to number of users.]{
    \label{fig:gdm1-user} 
    \includegraphics[width=5.7cm]{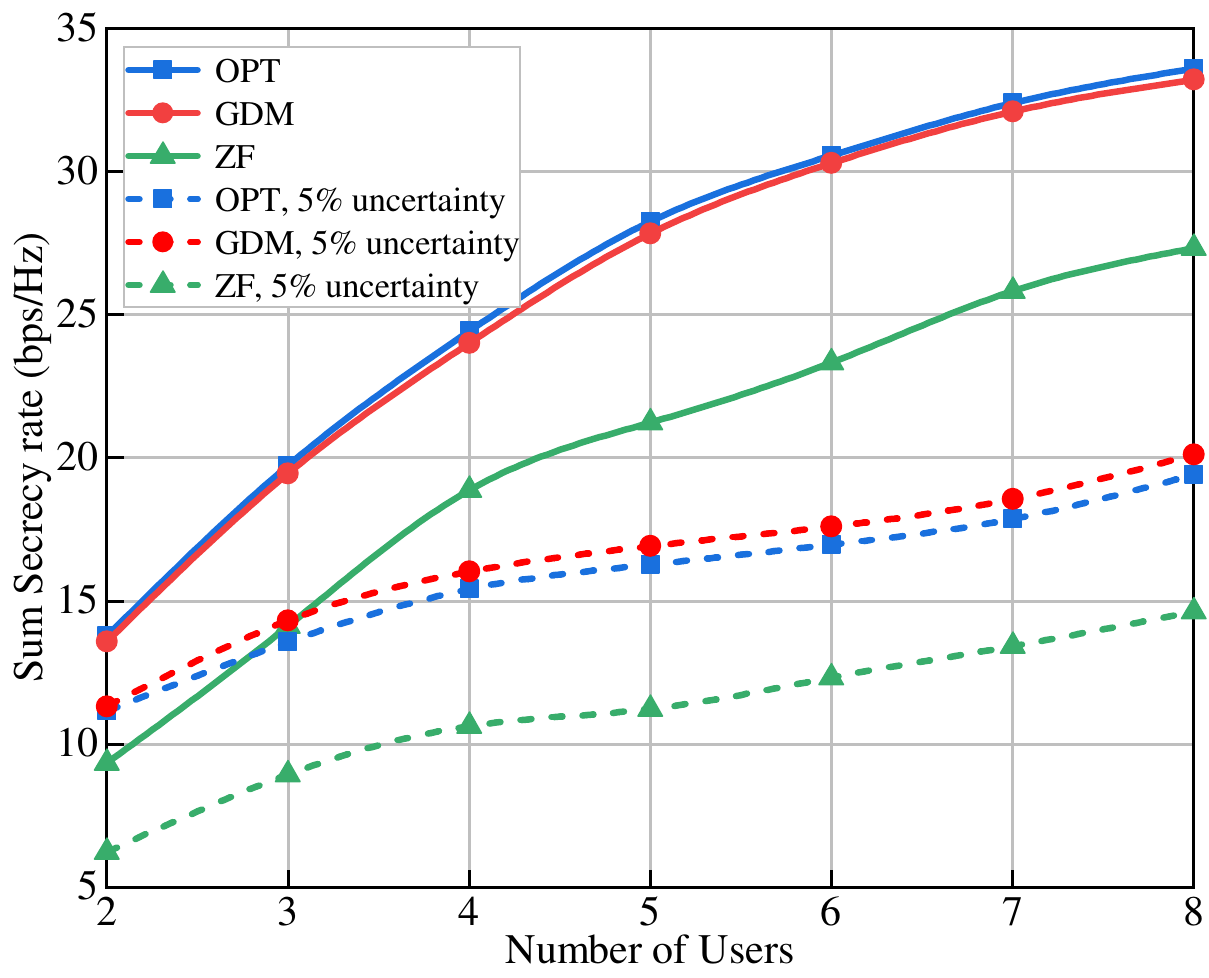}}
  \subfigure[Performance with respect to number of antennas.]{
    \label{fig:gdm1-ant} 
    \includegraphics[width=5.7cm]{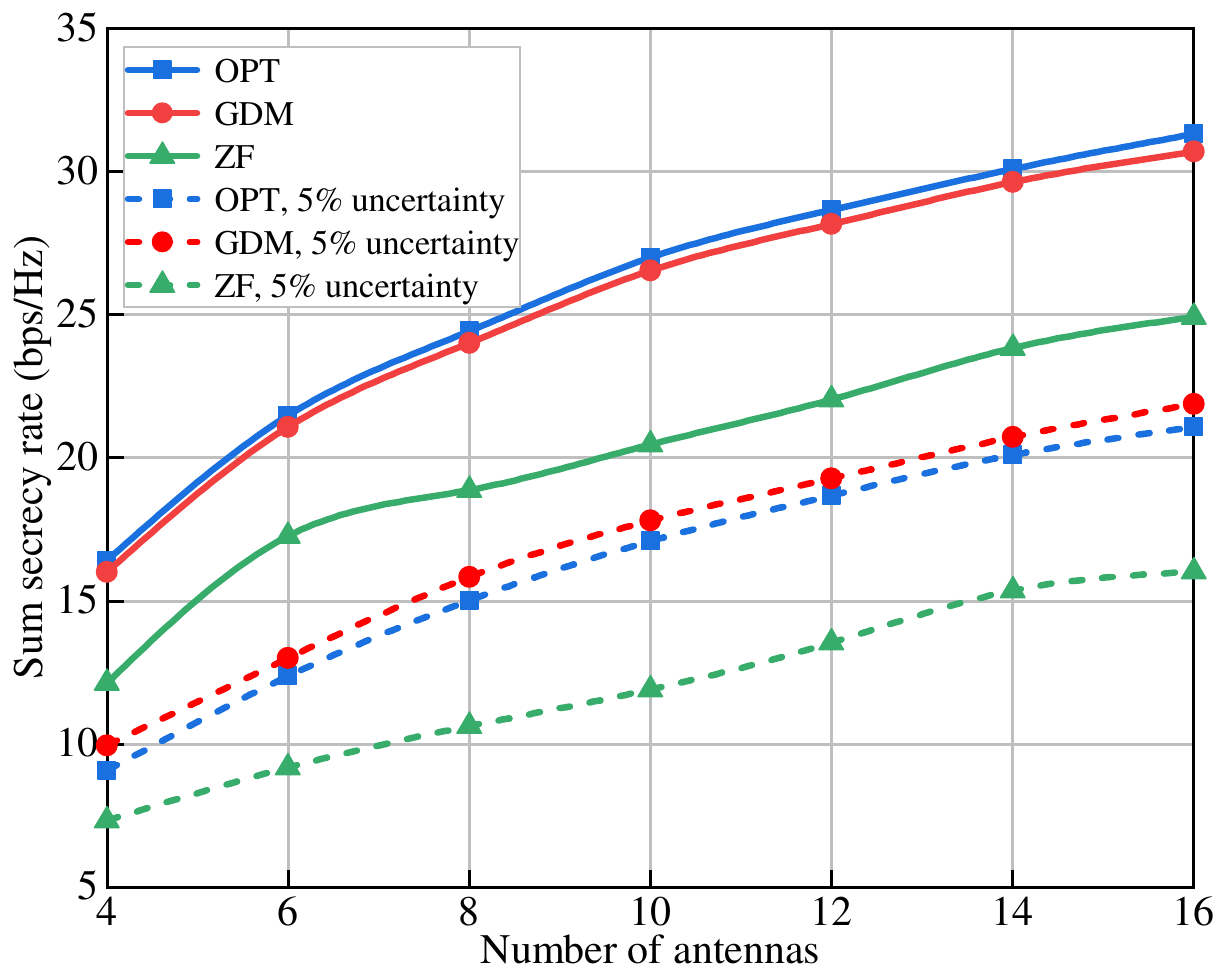}}
    \subfigure[Performance with respect to transmit power.]{
    \label{fig:gdm1-pwr} 
    \includegraphics[width=5.7cm]{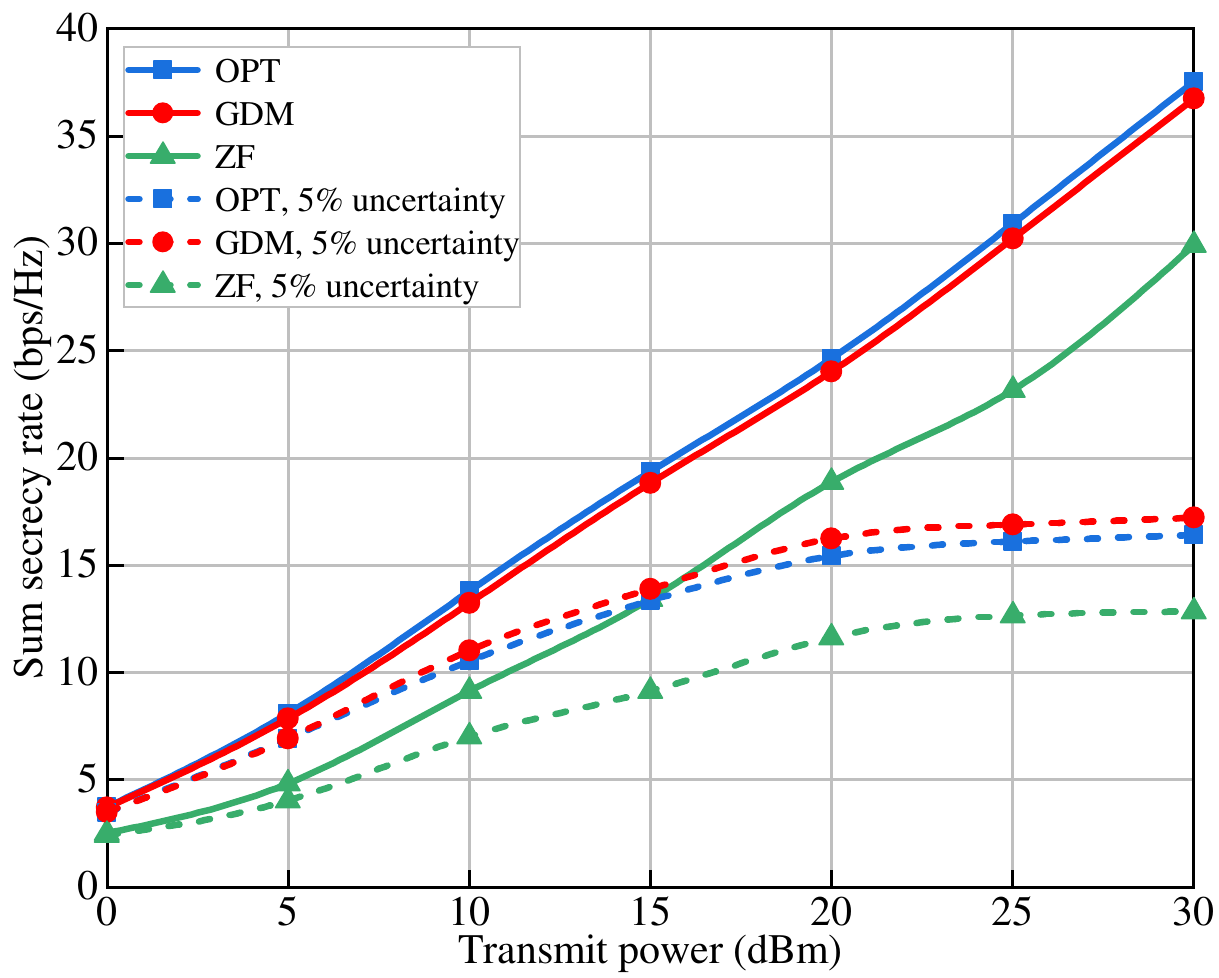}}
  }
  \caption{Performance of GDM expert for wireless scenario with $ \kappa_L=2 $, $ \kappa_R=10 $ (light attenuation with strong LoS, typically open-field rural).}
  \label{fig:gdm1} 
\end{figure*}

\begin{figure*}[t] 
  \centerline{
  \subfigure[Performance with respect to number of users.]{
    \label{fig:gdm2-user} 
    \includegraphics[width=5.7cm]{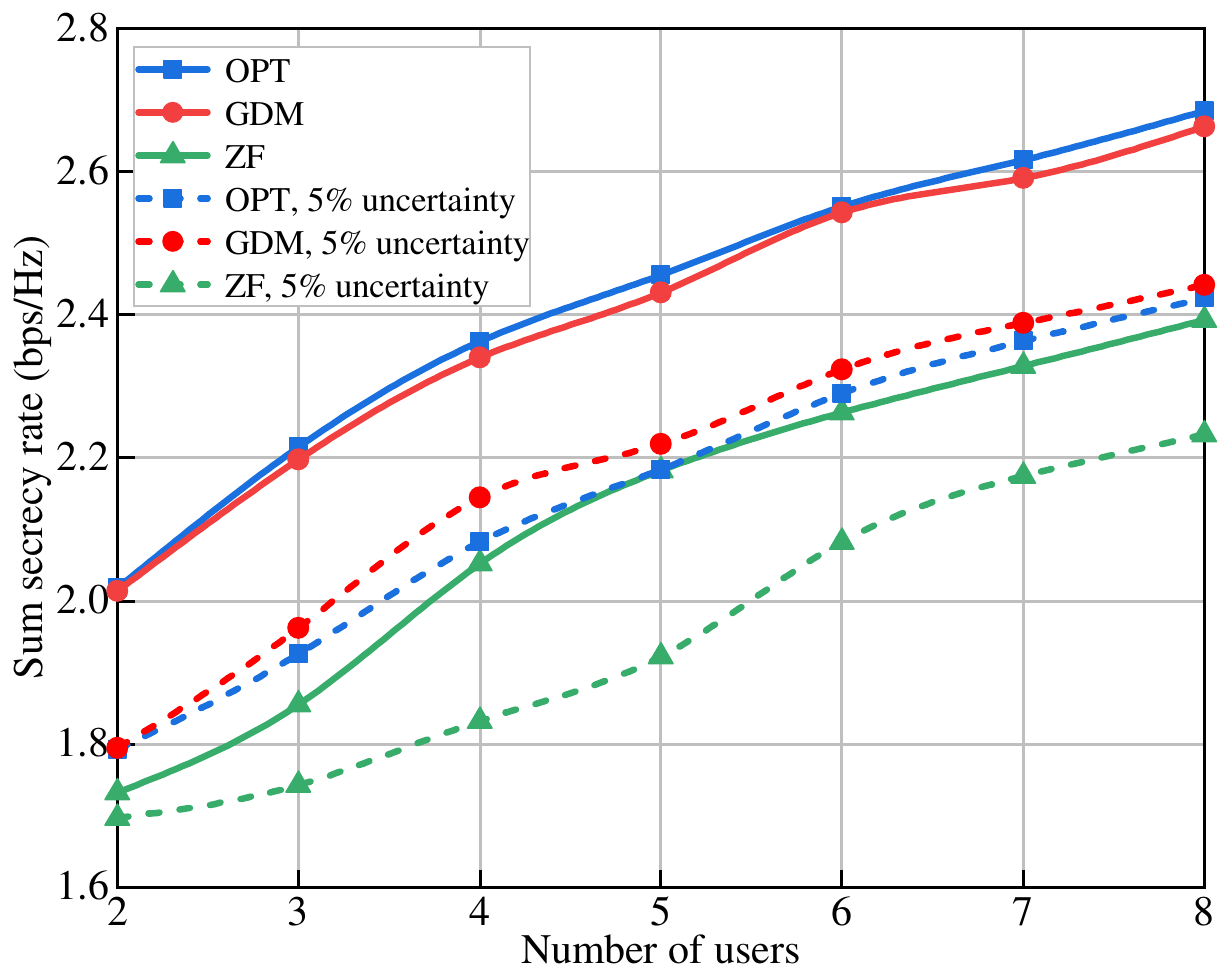}}
  \subfigure[Performance with respect to number of antennas.]{
    \label{fig:gdm2-ant} 
    \includegraphics[width=5.7cm]{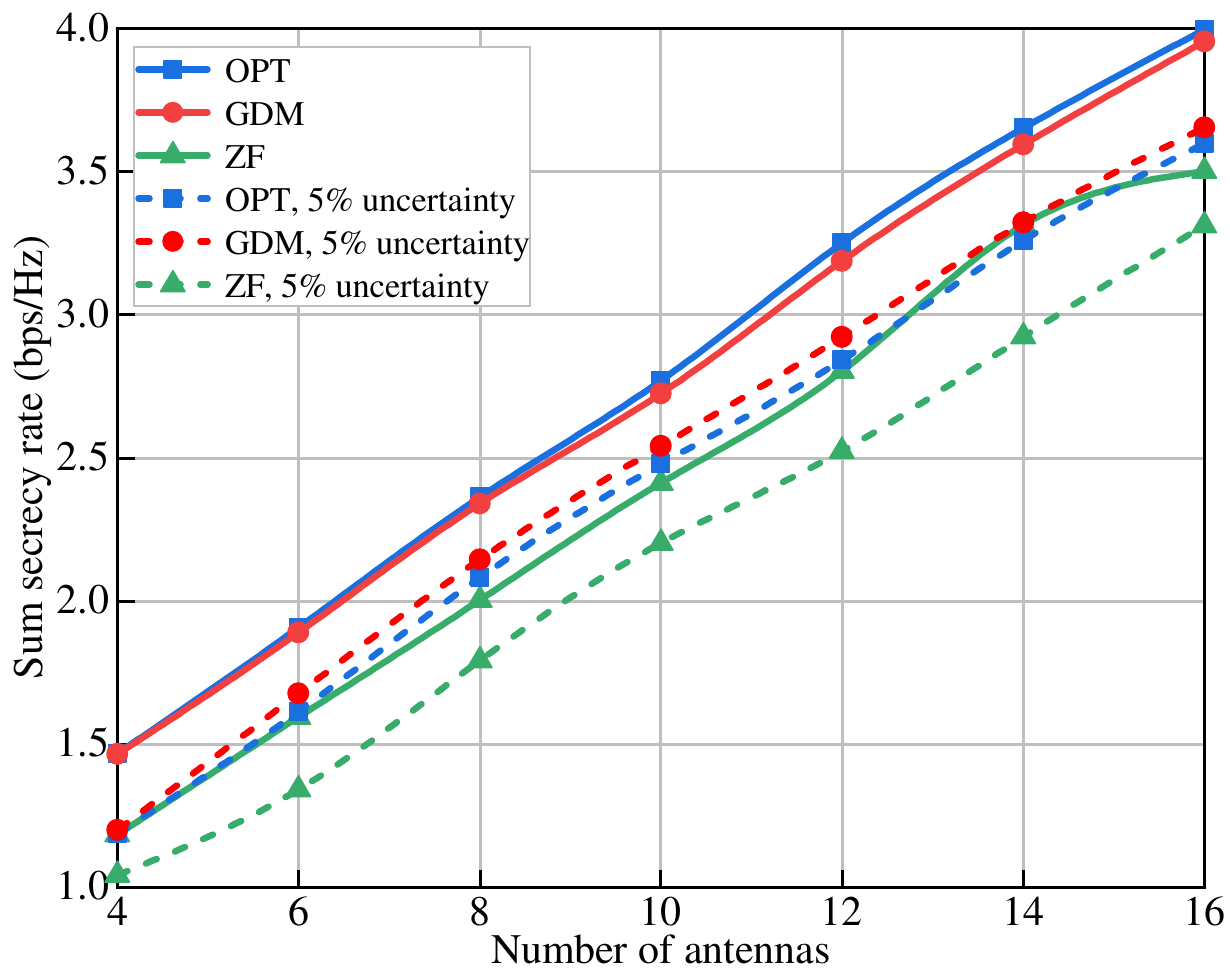}}
    \subfigure[Performance with respect to transmit power.]{
    \label{fig:gdm2-pwr} 
    \includegraphics[width=5.7cm]{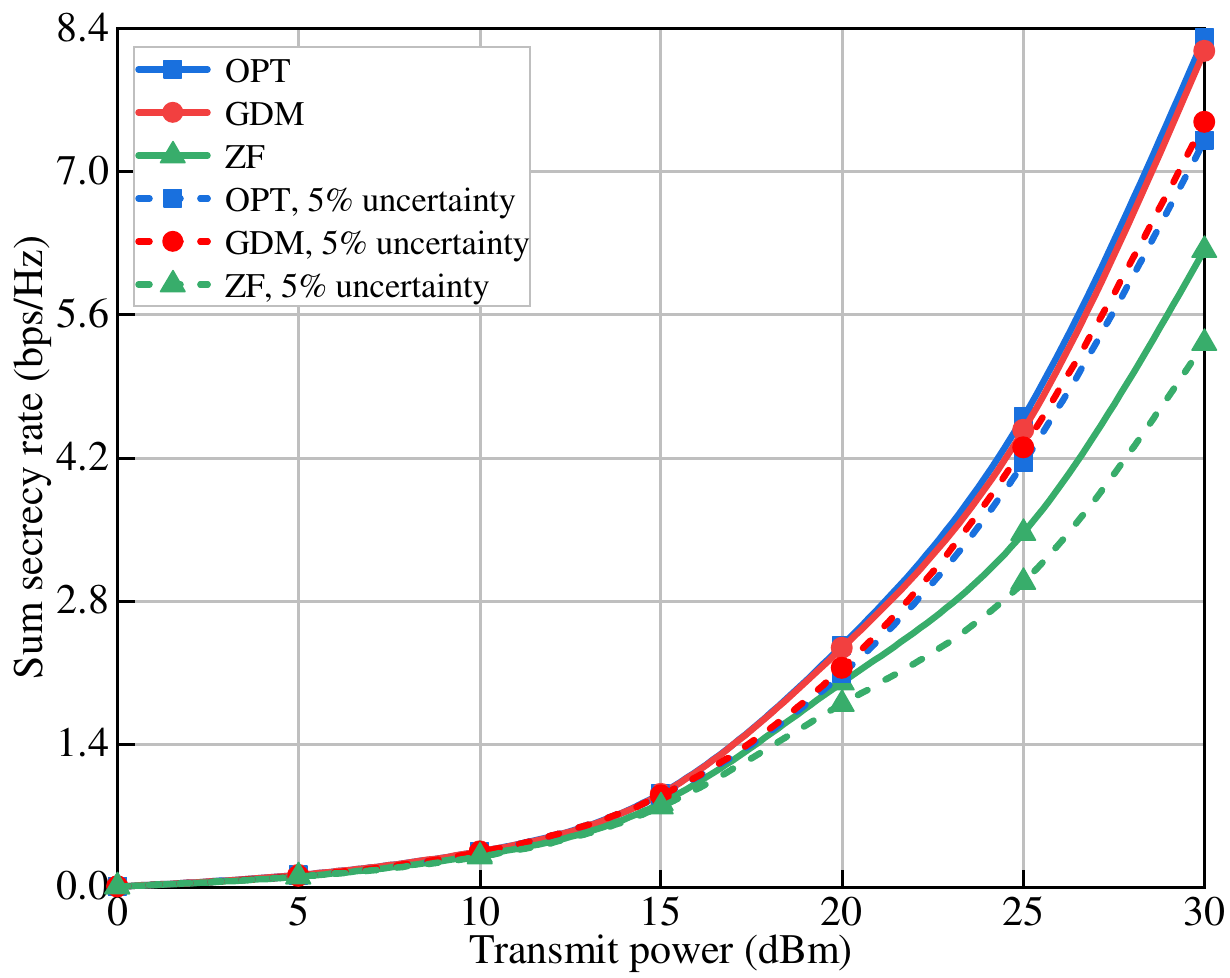}}
  }
  \caption{Performance of GDM expert for wireless scenario with $ \kappa_L=4 $, $ \kappa_R=1 $ (high attenuation with weak LoS, typically dense urban).}
  \label{fig:gdm2} 
\end{figure*}

\subsection{Performance of GDM Experts}

We show the performance of the single GDM experts, where for illustration purpose, we show two experts within the representative scenario set. In this respect, we evaluate the performance of experts and demonstrate the advantage of diffusion models. For performance comparison, we consider an optimization-based benchmark (OPT) and a conventional zero-forcing (ZF) secure-beamforming scheme. In this regard, ZF provides a lightweight conventional reference with a closed-form spatial structure, while OPT provides a computationally more demanding per-instance optimization reference that directly targets the secrecy-rate objective. Moreover, consider the practical channel state information issue, we also evaluate the performance when the channel matrices are associated with 5\% uncertainties.

Fig.~\ref{fig:gdm1} presents the performance of the GDM-based security expert trained for a specific wireless scenario characterized by light path loss and a strong Line-of-Sight (LoS) component ($\kappa_L=2, \kappa_R=10$). As illustrated across all three subfigures, the sum secrecy rate consistently improves with more system resources, such as a higher number of users, antennas, or greater transmit power. The key observation is the remarkable performance of our GDM expert, which operates nearly identically to the OPT benchmark under the assumption of perfect channel state information. This reveals that the GDM experts has the capability to learn the inherent structure of the optimal solution space. establishing the complex mapping from channel conditions to the optimal security beamforming strategy. Moreover, the GDM expert can be adaptive to the variations of network configurations of the considered factor, allowing significant generalization capability of the learning model.

More importantly, the results highlight the GDM's robustness in the presence of imperfect channel information. When a 5\% uncertainty is introduced (dashed lines), the GDM expert consistently outperforms the OPT method. This suggests that the GDM, through its training on diverse data, has learned a more robust strategy that is less susceptible to channel estimation errors compared to the optimization algorithm. The ZF method, experiences the most significant performance degradation under uncertainty. This makes the GDM model particularly applicable for practical wireless systems, as the channel information is inevitably associated with errors while the GDM-based design can more effectively maintain the reliable security performance.

Fig.~\ref{fig:gdm2} evaluates the GDM expert specifically trained for an environment characterized by high path loss and a weak LoS component ($\kappa_L=4, \kappa_R=1$). The performance trends are consistent with the previous scenario, showing that the sum secrecy rate scales positively with the number of users, antennas, and transmit power, but the overall achieved secrecy rate is much lower due to the aggravated environment. Particularly, in this more challenging propagation environment, our GDM expert continues to demonstrate strong performance, closely approaching the optimization benchmark and dominating the ZF baseline. This confirms the model's capability to learn effective security strategies even in harsh conditions.

When channel uncertainty is introduced, the GDM expert remains robust regardless of the uncertainties. Similarly in this harsh scenario, the GDM consistently outperforms the optimization benchmark under uncertainty, suggesting that the benefits of its learned robustness are reliable for different scenarios. The performance gap between the GDM and the optimal solution, especially when contrasted with the simpler ZF baseline, underscores the effectiveness and reliability of our proposed approach across different and challenging wireless environments.

\subsection{Performance of the Router}

\begin{figure}[t]
   \centering
   \includegraphics[width=0.44\textwidth]{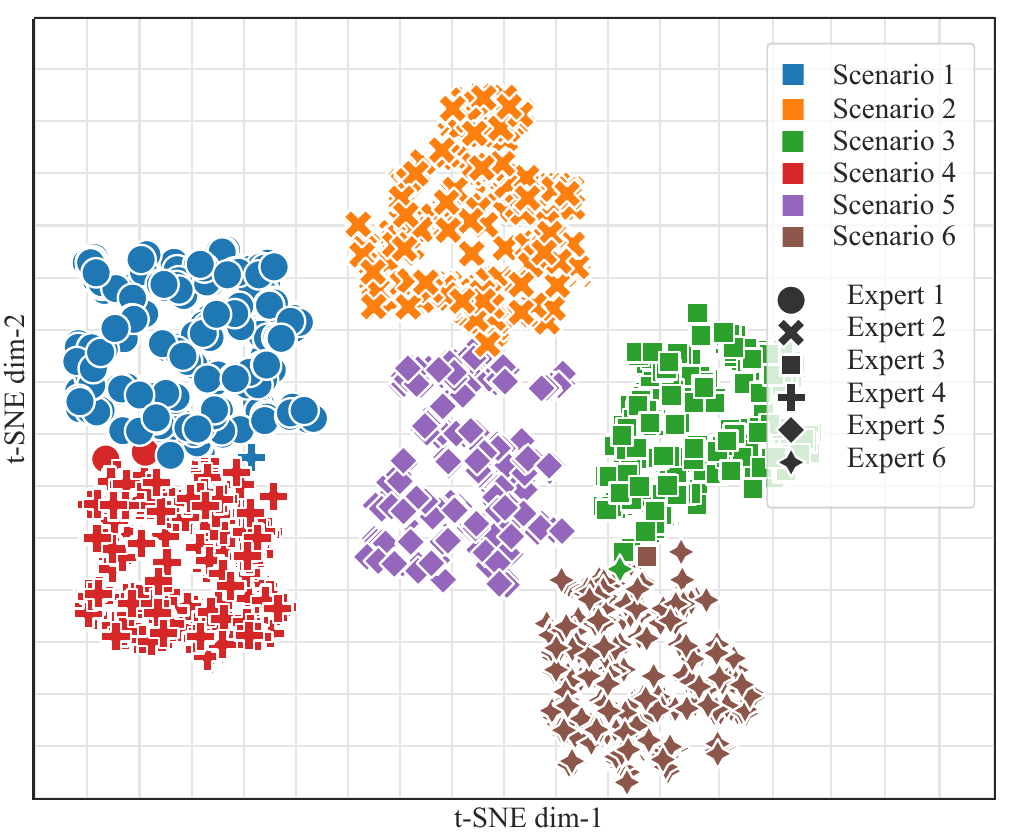} 
   \caption{The t-SNE visualization of the engineered channel features for the 6 representative scenarios.}
   \label{fig:tsne}
\end{figure}

\begin{figure}[t]
   \centering
   \includegraphics[width=0.45\textwidth]{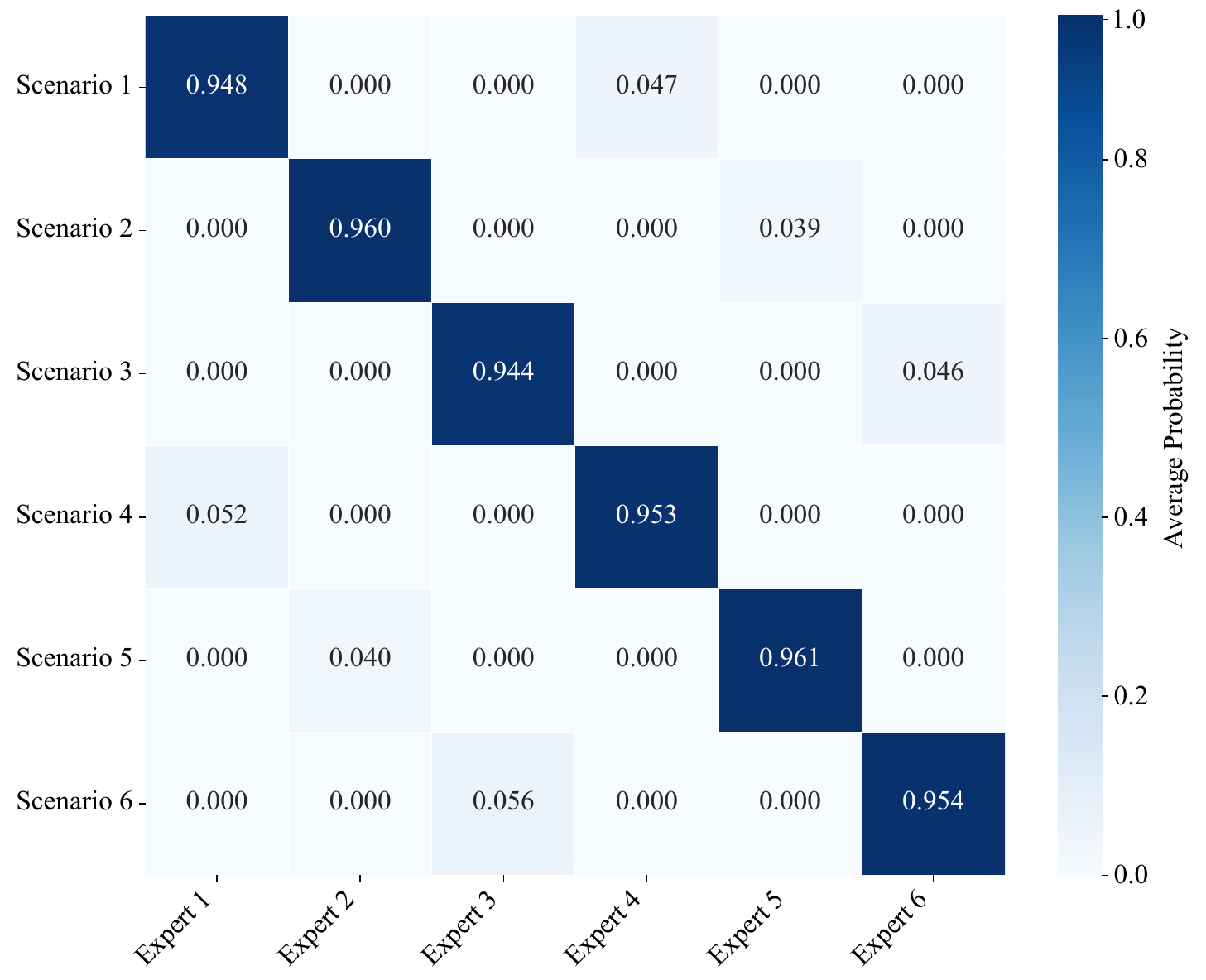} 
   \caption{Heatmap of the router with average expert selection probabilities.}
   \label{fig:ht}
\end{figure}

To evaluate the efficacy of the proposed MoE architecture, we first present a validation of the router, or the gating network, which serves as the core component for scenario identification and expert orchestration. The performance of the router is visualized in Fig.~\ref{fig:tsne} and Fig.~\ref{fig:ht}, which respectively demonstrate the separability of the engineered channel features and the classification accuracy of the router.

Fig.~\ref{fig:tsne} provides a t-SNE visualization that projects the 6-dimensional engineered feature vectors, extracted from the channel matrices under the 6 representative scenarios, into a two-dimensional space for qualitative assessment. The results show 6 clearly delineated and well-separated clusters, where each cluster corresponds to one of the predefined wireless scenarios. This distinct clustering indicates that the proposed feature engineering is rather effective to captures the statistical properties of each scenario. This established separation is the foundation towards the scenario identification and expert association, which is further used to support the combination of expert output for final security strategy determination.

Meanwhile, Fig.~\ref{fig:ht} shows the performance of the router through a heatmap of the average expert selection probabilities. This matrix illustrates the probability of the gating network to select a particular expert (columns) when presented with input from a specific ground-truth scenario (rows). The matrix exhibits strong diagonal dominance, with probabilities for correct expert selection always exceeding 0.94. The near-zero values in the off-diagonal elements indicate a negligible rate of misclassification. The miss classified cases that happen rarely are also demonstrated in Fig.~\ref{fig:tsne} with the misplaced markers and color identifications. As we can see in Fig.~\ref{fig:tsne}, the miss classifications are only encountered at the small intersections of different clusters. Nevertheless, the overall high classification accuracy implies that the router can reliably map the observed channel environment to the specialized experts, supporting the MoE framework to cover diverse scenarios.

\subsection{Cross-Scenario Security Performance}

\begin{figure*}[t]  
  \centerline{
  \subfigure[Performance under representative scenarios.]{
    \label{fig:moe_1} 
    \includegraphics[width=7.5cm]{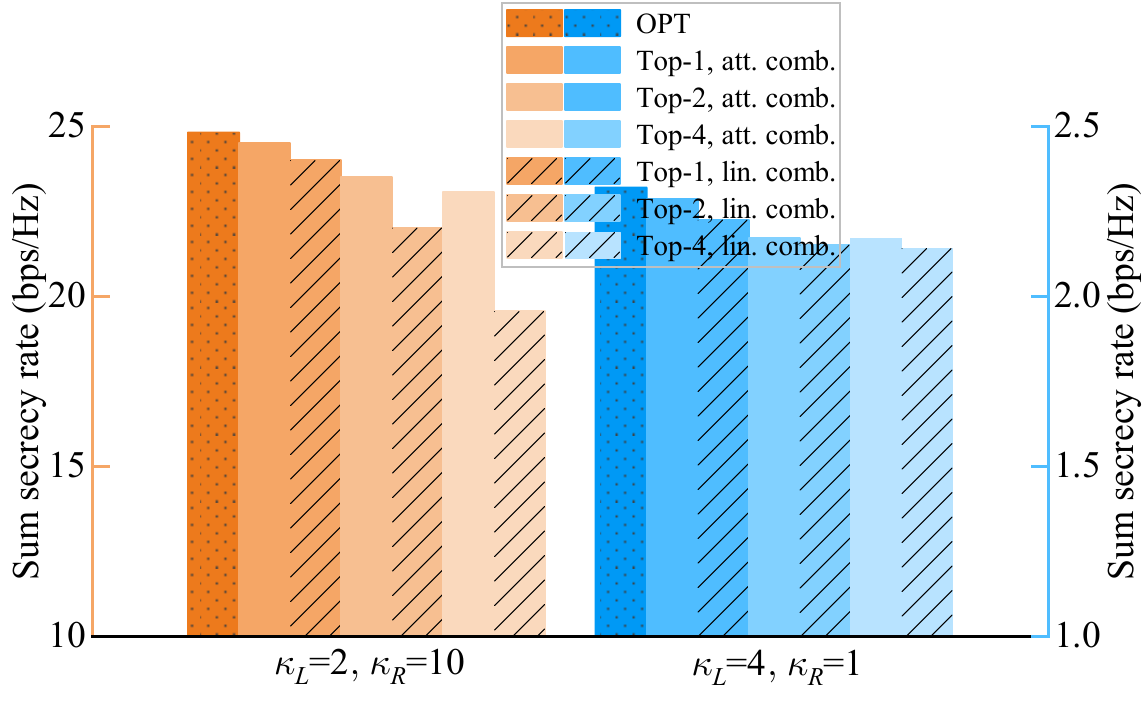}}
  \quad
  \subfigure[Performance under cross-scenarios (interpolated Rician factors).]{
    \label{fig:moe_2} 
    \includegraphics[width=7.5cm]{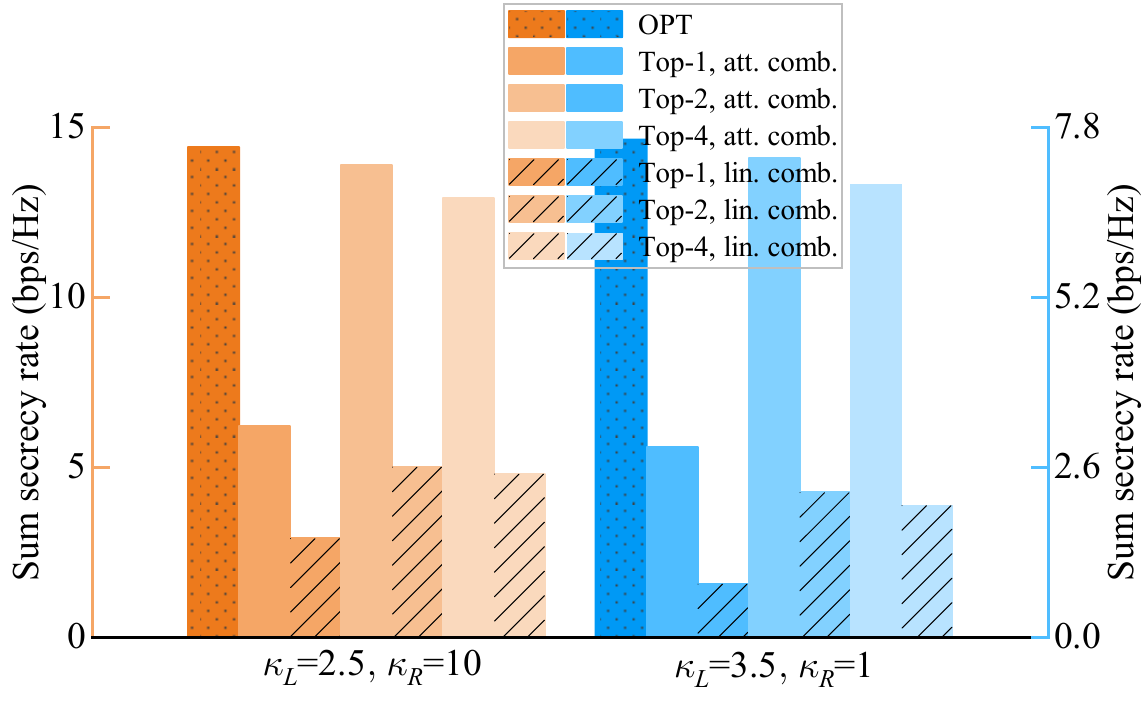}}
   }
    \centerline{
  \subfigure[Performance under cross-scenarios (interpolated path loss exponents).]{
    \label{fig:moe_3} 
    \includegraphics[width=7.5cm]{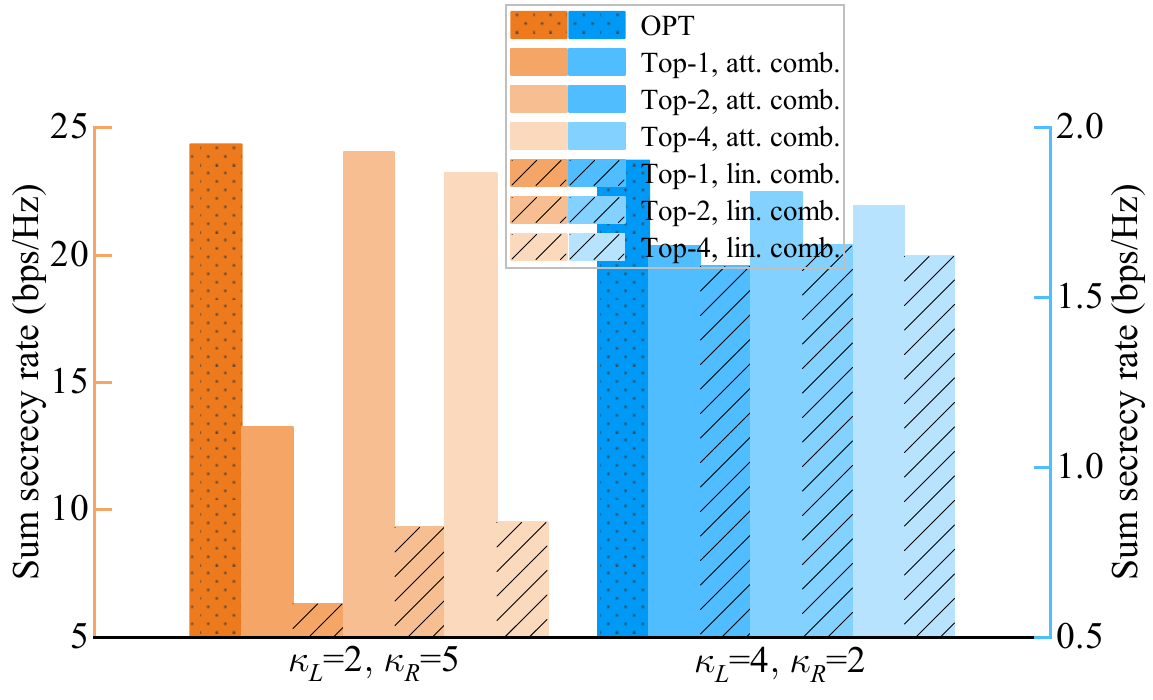}}
  \quad
  \subfigure[Performance under cross-scenarios (interpolated path loss exponents and Rician factors).]{
    \label{fig:moe_4} 
    \includegraphics[width=7.5cm]{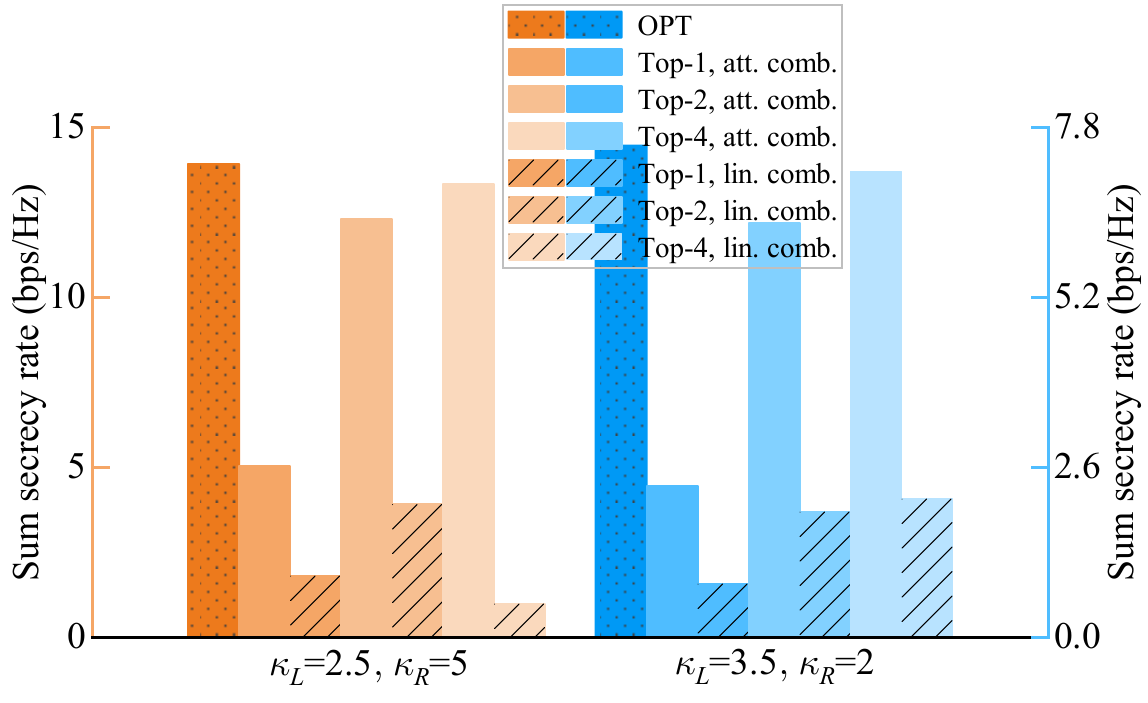}}
  }
  \caption{Performance comparison under diverse wireless scenarios.}
  \label{fig:moe-all} 
\end{figure*}

We now evaluate the end-to-end performance of the complete MoE framework, with the results presented in Fig.~\ref{fig:moe-all}. This evaluation is designed to corroborate the core design goal of our architecture to effectively recognize and cover the unseen cross-scenarios. This requires, first, that the gating network can effectively select the appropriate expert for a given environment, and second, that the attention-based combiner can intelligently synthesize the expert output to achieve robust security across a wide varieties of unseen wireless scenarios. The results demonstrate the performance of using the top-1, top-2, and top-4 expert selection, both with our attention-based combiner and with a simple weighted-average combination baseline.

Fig.~\ref{fig:moe_1} presents the performance in the two representative scenarios upon which experts were explicitly trained. In these cases, the highest secrecy rate is achieved when only the single, top-rated expert is selected (Top-1). This result serves as a crucial sanity check, confirming that the gating network correctly identifies the single best expert for the known environments. When additional experts are included (Top-2, Top-4), the security performance can only get downgraded due to the intervention from inappropriate experts. Therefore, the model effectiveness is not necessarily monotonic with the number of selected experts, because expanded expert set may include less relevant expert and also increase the synthesis difficulty for the combiner. As a further note, we also show the results obtained through optimization, which is in consistency with those in Figs.~\ref{fig:gdm1} and~\ref{fig:gdm2} that the top-1 expert approximates the optimum.

The adaptability and robustness of the MoE architecture become more evident in the cross-scenarios as shown in Fig.~\ref{fig:moe_2}, \ref{fig:moe_3}, and \ref{fig:moe_4}. These figures test the system in unseen environments that interpolate the characteristics of the representative scenarios. In contrast to the previous case, relying on a single expert (Top-1) here leads to certain inferior performance. The optimal strategy in these interpolated environments is not captured by any single expert but by activating multiple relevant experts (Top-2 and Top-4) to provide the candidate solutions to achieve a high secrecy rate. Moreover, to further quantify the contribution of the combiner module, the optimization-based reference is included. Evidently, the results show that the attention-based residual synthesis consistently approaches the optima, verifying that the proposed combiner effectively learns the non-linear mapping beyond linear interpolation. This observation is particularly pronounced when multiple experts are activated under the unseen cross scenarios, where the gap of weighted linear average to optima becomes evident, while the attention-based combiner reliably maintains near-optimal performance.

Therefore, the cross-scenario results verifies the reliable security guarantee in unknown environments, which relies on the proper expert selection by the router and the effective synthesis by the attention-based combiner. The approach of simply weighted averaging the outputs of the selected experts consistently yields poor results, as linear interpolation is insufficient to approximate the complex, non-linear solution space of secure beamforming. In contrast, the attention-based combiner intelligently weighs the expert proposals and computes a residual correction, transforming the simple baseline into a high-fidelity security strategy. In this regard, the performance gain can be substantial, particularly when the scenario becomes more complex and more ``cross'' and thus requires the collaboration from more experts. Overall, these results validate the MoE framework can effectively orchestrate specialized knowledge to provide robust and superior security across diverse wireless landscapes.

As a further note, the current cross-scenario evaluation mainly considers propagation regimes within the coverage of the representative scenario set. Generalization to strongly extrapolated or fundamentally different channel environments is more challenging and is not guaranteed by the present framework, and extending the expert coverage and developing more explicit out-of-distribution adaptation mechanisms constitute important future directions. Additionally, the present evaluation focuses on the average sum secrecy rate in accordance with the secrecy-rate maximization objective. Nevertheless, the proposed framework can be conveniently extended to cover metrics like variance, secrecy outage, and lower-tail performance without changing the basic GDM-based MoE architecture.

\section{Conclusion} \label{sec:con}

In this paper, we have proposed a novel and adaptive framework to address the challenge of physical layer security provisioning across diverse wireless scenarios. We have proposed a MoE architecture that orchestrates a committee of specialized security experts, and each expert is realized as a powerful GDM with a Transformer-based denoising network. We also have designed a gating network for expert selection and a combination network to intelligently synthesizes the expert outputs for security strategy determination. We have showed that the MoE framework effectively adapts to unseen environments, providing reliable, robust, and superior cross-scenario physical layer security for future 6G networks.

\bibliographystyle{IEEEtran}
\bibliography{main}

\end{document}